\documentclass[aps,prd,10pt,nofootinbib,notitlepage,superscriptaddress,
tightenlines,floats,floatfix]{revtex4}

\usepackage{amsmath}  
\usepackage{amsfonts}
\usepackage{amssymb}
\usepackage{graphics}
\usepackage{graphicx}
\usepackage{epsf}
\usepackage{xspace}
\usepackage{color}
\usepackage{url}
\usepackage{fnbreak}
\usepackage{float}

\usepackage[normalem]{ulem}
\usepackage{color}

\def\be{\begin{equation}}
\def\ee{\end{equation}}
\def\bea{\begin{eqnarray}}
\def\eea{\end{eqnarray}}
\def\nn{\begin{nonumber}}
\begin{document}

%%%%%%%%%%%%%%%%%%%%%%%%%%%

% include preprint number option
%\preprint{{\br  Version 11}} %   LFTC-23-5/78}}

%\vspace{-1in}%\parbox{1.5in}{ \vspace{-9.6in}}  % moves the preprint box down

\vspace*{-1.2cm}
\title{
%\vspace{-1.5cm}
  \begin{flushright}
    LFTC-23-10/83
  \end{flushright}
  \vspace{0.5cm}
In-medium mass shift of two-flavored heavy mesons, \boldmath{$B_c$}, \boldmath{$B^*_c$},
\boldmath{$B_s$}, \boldmath{$B^*_s$}, \boldmath{$D_s$} and \boldmath{$D^*_s$}}
\author{G.~N.~Zeminiani}
\email{guilherme.zeminiani@gmail.com}
\affiliation{Laborat\'orio de F\'isica Te\'orica e Computacional (LFTC),
Universidade Cidade de S\~ao Paulo (UNICID), 
01506-000, S\~ao Paulo, SP, Brazil}

\author{S.~L.~P.~G.~Beres}
\email{samuel.beres@hotmail.com}
\affiliation{Laborat\'orio de F\'isica Te\'orica e Computacional (LFTC),
Universidade Cidade de S\~ao Paulo (UNICID), 
01506-000, S\~ao Paulo, SP, Brazil}

\author{K.~Tsushima}
\email{kazuo.tsushima@gmail.com,
kazuo.tsushima@cruzeirodosul.edu.br}
\affiliation{Laborat\'orio de F\'isica Te\'orica e Computacional (LFTC),
Universidade Cidade de S\~ao Paulo (UNICID),  
01506-000, S\~ao Paulo, SP, Brazil}
%%%%%%%%%%%%%%%%%%%%%%%%%%%%%%

%G.~Zeminiani
%\\
%\email{guilherme.zeminiani@gmail.com}
%Laborat\'orio de F\'isica Te\'orica e Computacional, Universidade Cruzeiro 
%do Sul / Universidade Cidade de Sao Paulo, 01506-000, S\~ao Paulo, SP, Brazil
%\\

\date{\today}

%\maketitle

%%%%%%%%%%%%%%%%%%%%%%%%%%%%%%%%%%%%%%%%%%%%%%%%%%
%\section{\leftline{abstruct \hfill}}
%\vspace{-3ex}
%%%%%%%%%%%%%%%%%%%%%%%%%%%%%%%%%%%%%%%%%%%%%%%%%%

%%%%%%%%%%%%%%%%%
\begin{abstract}
%%%%%%%%%%%%%%%%%
For the first time, we estimate the in-medium mass shift of
the two-flavored heavy mesons $B_c, B_c^*, B_s, B_s^*, D_s$ and $D_s^*$ in symmetric nuclear
matter. The estimates are made by evaluating the lowest order one-loop self-energies.
The enhanced excitations of intermediate state
heavy-light mesons in symmetric nuclear matter are the origin of their negative mass shift.
This negative mass shift may be regarded as a signature of
partial restoration of chiral symmetry
in an empirical sense 
because the origin of the negative mass shift in the study
is not directly related to the chiral symmetry mechanism.
Our results show that the magnitude of the mass shift for the $B_c$ meson ($\bar{b} c$ or $b \bar{c}$)
is larger than
those of the $\eta_c (\bar{c} c)$ and $\eta_b (\bar{b} b)$,
different from a naive expectation that it would
be in between them.
While, that of the $B_c^*$ shows the in between of the $J/\psi$ and $\Upsilon$.
We observe that the lighter vector meson excitation
in each meson self-energy gives a dominant contribution for the corresponding meson mass shift,
$B_c, B_s,$ and $D_s$.

%%%%%%%%%%%%%%
\end{abstract}
%%%%%%%%%%%%%%

\maketitle

%%%%%%%%%%%%%%%%%%%%%%%%%%%%%%
\section{Introduction}
%%%%%%%%%%%%%%%%%%%%%%%%%%%%%%

The study of hadronic interactions with the nuclear medium,
particularly for those composed of only heavy quarks,
is one of the excellent ways to explore the roles of gluons
based on quantum chromodynamics (QCD),
because such hadrons, like two-heavy flavored mesons,
do not share light quarks with the nucleons, and thus,
their interactions with the nuclear medium
are expected to occur primarily mediated by gluons.

The production of heavy mesons in high energy heavy ion collisions such as
in RHIC~\cite{Schroedter:2000ek} and
the LHC~\cite{Andronic:2015wma}, is also a major motivation
for understanding hadron-nucleus (nuclear medium) interactions.
In such cases, the medium modifications of hadron properties
can have a significant impact on the experimental results.
Recently, huge interest has been drawn upon the properties
of two-heavy flavored mesons, in particular the $B_c$ meson
family~\cite{Li:2023wgq,Arnaldi:2023zlh,Borysova:2022nyr,Liu:2022bdq,CMS:2022sxl,Liu:2023kxr}.
Although the information for the $B_c$ meson family remains limited compared to that for quarkonia~\cite{Li:2023wgq},
the $B_c$ ($B_c^*$) meson may naively be expected to
have intermediate properties between those of the charmonium
$\eta_c$ ($J/\psi$) and bottomonium $\eta_b$ ($\Upsilon$)~\cite{Lodhi:2007zz},
although the $B^*_c$ meson existence has not yet been confirmed experimentally~\cite{Liu:2022bdq}.
Focusing on the point as to how $B^*_c$ meson can be
identified experimentally, a theoretical study was made recently~\cite{Geng:2023ffc}.

In the early 1990s, it was proposed that the $B_c$ meson could potentially be detected
in hadron colliders~\cite{Chang:1992bb,Chang:1992jb}.
Many attempts to observe the $B_c$ meson were made
led by various groups, such as the DELPHI~\cite{DELPHI:1996vyn},
ALEPH~\cite{ALEPH:1997oob}, and OPAL~\cite{OPAL:1998gdf} Collaborations.
It was in 1998 that the CDF collaboration finally observed a $B_c$ meson in
a $p\overline{p}$ collision via the decay mode
$B^{\pm}_c \rightarrow J/\Psi l^{\pm} \nu$~\cite{CDF:1998ihx}.
Furthermore, in 2014, the ATLAS Collaboration observed the excited state
$B_c$(2S)~\cite{ATLAS:2014lga}.
Recently, a study of $B_c$ meson state family was extensively
made~\cite{Li:2022bre,Li:2023wgq} using the Cornell potential~\cite{Eichten:1978tg}.

Our interest for the $B_c$ meson lies in
its properties in a dense nuclear medium, which can be of great relevance
for neutron star and magnetar structure, like the charmed mesons studied
in Ref.~\cite{De:2023vsb}.
Moreover, some studies have suggested that the $B_c$ meson
can serve as a probe of quark-gluon plasma
(QGP)~\cite{Akram:2013dhd,Lodhi:2007zz,Lodhi:2011zz,Wu:2023djn,CMS:2022sxl}.

The $J/\Psi$ suppression is often regarded as a signal
of QGP formation~\cite{Akram:2013dhd,Lodhi:2007zz,Lodhi:2011zz} 
since one of the possible dissociation 
mechanisms is the Debye screening of color interaction 
that would occur in a deconfined phase.
However, the suppression can also occur due to
the hadron interactions, such as absorption
by the medium modified comoving hadrons~\cite{Sibirtsev:1999jr,Lodhi:2011zz}.
The $\Upsilon$ meson is also expected to melt in QGP as a result of the color screening,
but due to the different binding energies of the $b\overline{b}$ states
from those of the $c\overline{c}$,
it would occur at different temperatures in QGP~\cite{Harris:2023tti} from
the $c\bar{c}$ states. It was suggested that the $\Upsilon$ absorption by hadronic
comovers may be insignificant~\cite{Lin:2000ke}, but still remains to be
understood better whether or not the observed
suppression~\cite{CMS:2018zza,ATLAS:2022exb,STAR:2022rpk} comes from QGP.
In QGP, the production rate of two-heavy flavored hadrons
would also be affected~\cite{Lodhi:2011zz}, but opposite to the heavy quarkonia,
the production of the $B_c$ meson could be enhanced in QGP.
This is due to the abundance of unpaired $b$ ($\overline{b}$)
and $c$ ($\overline{c}$) quarks in QGP, as the result of
plenty amount of melting quarkonia. Those $b$ and $c$ quarks
($\overline{b}$ and $\overline{c}$ antiquarks)
will be able to form large amount of
$B_c$ mesons~\cite{Akram:2013dhd,Lodhi:2007zz,Lodhi:2011zz}.
In addition, $B_c$ is more stable than quarkonia,
since the difference in quark flavors avoids the quark and antiquark annihilation,
and is expected to decay weakly.
(However, there are some theoretical predictions/studies
for the $B_c$ meson strong decays~\cite{Li:2023wgq}.)
The reasons given above can make the $B_c$ meson as an
interesting probe of QGP.

A possible mechanism for the two-heavy flavored meson interaction with
the nuclear medium, aside from those arising from gluonic origins, is through the excitation of
the intermediate state hadrons with light quarks.
This mechanism was explored in the study of heavy charm and bottom
quarkonia~\cite{Krein:2010vp,Krein:2017usp,Zeminiani:2020aho,Cobos-Martinez:2020ynh}.
We extend the approach applied for the heavy quarkonia to the two-heavy flavor mesons
in this study. In particular, our focus is on the in-medium mass shift of the bottom-charmed
$B_c$ and $B^*_c$ mesons.
This will provide us with the opportunity to compare the strengths of meson-nuclear matter
interactions among quarkonia and two-heavy flavor mesons.
Another interest is that, whether or not the naive expectation that the
(bottom-charmed meson)-(nuclear matter) interaction strength
is in between those of the (charmonia-nuclear matter) and (bottomonia-nuclear matter).

Furthermore, we study the heavy-strange mesons,
$B_s$, $B^*_s$, $D_s$ and $D^*_s$.
These mesons may show some characteristic feature in the mass shift
compared with those of the charmonia, bottomonia, $B_c$, and $B_{c}^*$.
Such heavy-strange mesons are also possible probes
of QGP~\cite{ALICE:2022xrg}. ($B^0_s$ properties in free space were
studied in Refs~\cite{Das:2019cpt,Das:2021lws,Pandya:2023ldv}.)

The transition $b \rightarrow s \ell^+ \ell^-$ can trigger decays of the type
$B_c \rightarrow D_s \ell^+ \ell^-$ and 
$B_c \rightarrow D^*_s \ell^+ \ell^-$~\cite{Li:2023mrj}.
This flavor-changing neutral-current process is highly suppressed in the Standard Model (SM), thus
opening a way for new physics beyond the SM, which attracted significant
attention for $B_c$ meson~\cite{Bird:2004ts,Altmannshofer:2014rta,Buras:2014fpa,
Descotes-Genon:2015uva,Dutta:2017xmj,CMS:2021hug}.
However, the $B_c \rightarrow D_s \mu^+ \mu^-$ process has not yet been observed.
A recent search for this rare decay led by
the LHCb Collaboration has found no significant signals
of this process in the nonresonant $\mu^+ \mu^-$ modes~\cite{LHCb:2023lyb}.
Furthermore, semileptonic decays of $B_c$ into excited states
of $D_s$ and $B_s$ have also been studied~\cite{Hazra:2023zno} and similarly the 
decay of $B_s$ into excitations of $D_s$~\cite{Alonso-Alvarez:2023mgc,Penalva:2023snz}.
The decay rates of these processes should be modified in a nuclear medium if the mass shift of the
mesons
occur.

The mass shift of the mesons mentioned above
are estimated by calculating the self-energies with the excitations of intermediate state mesons
with light quarks $u$ or $d$.
We employ an SU(5) effective Lagrangian density with
minimal substitutions to get the interaction Lagrangians
for calculating the self-energies of two-heavy flavored mesons in free space as well as in
symmetric nuclear matter.
The necessary inputs for calculating the in-medium self-energy are the effective mass
(Lorentz-scalar) of the intermediate state mesons (the Lorentz-vector potentials cancel between the
excited mesons appearing in the self-energy loop, and no need in the present study).
The in-medium masses of the excited mesons appearing in the self-energy processes are calculated by
the quark-meson coupling (QMC) model invented by Guichon~\cite{Guichon:1987jp},
which has been applied for various studies
successfully~\cite{Guichon:1995ue,Tsushima:2002cc,Tsushima:1997df,Saito:1996sf,
Saito:2005rv,Tsushima:2019wmq,Krein:2010vp,Krein:2017usp}.

In the QMC model, the light quark coupling to the Lorentz-scalar-isoscalar $\sigma$ field
results in a nonlinear dependence of the hadron effective mass on the $\sigma$ field,
and as a consequence, the reduction of the hadron effective mass with increasing
nuclear density.
This mass reduction in medium may be associated with a signature of
partial restoration of chiral symmetry in an empirical sense, because
the QMC model itself does not explicitly manifest a mechanism
related with chiral symmetry.

We analyze the mass shift of the mesons in detail
by comparing each loop contribution in the self-energy.
The results of the $B_c$ meson are compared
with those of the $\eta_b$ and $\eta_c$,
and the results of $B^*_c$ are compared with those of
the $\Upsilon$ and $J/\psi$.
Different from our naive expectation, it turns out
that the $B_c$ mass shift amount is larger than those of the $\eta_b$ and $\eta_c$,
and not in between them. While, the mass shift amount of $B^*_c$ is in between the $\Upsilon$
and $J/\psi$.
Furthermore, we find that the lighter vector meson excitation in each self-energy
plays the dominant role for the corresponding $B_c, B_s$, and $D_s$ mass shift.

This article is organized as follows.
In Sec.~\ref{secqmc}, we discuss the $B$, $B^*$, $D$, $D^*$, $K$, and $K^*$ meson
effective Lorentz-scalar masses in symmetric nuclear matter calculated by the QMC model.
In Sec.~\ref{secmshft} we first study the mass shift of $B_c$ and $B^*_c$ mesons
in symmetric nuclear matter.
We present the effective Lagrangian densities for calculating the interactions
$B_c B^* D$, $B_c B D^*$ and $B^*_c BD$,
obtained from an SU(5) symmetric Lagrangian.
The obtained effective Lagrangians are used to estimate the self-energy
loop contributions.
After a detailed analysis of the mass shift of $B_c$ and $B_c^*$ mesons,
we compare the results with those of the low-lying
heavy quarkonia, $\overline{b}b$ and $\overline{c}c$.
Furthermore, we estimate the mass shift of
the heavy-strange $B_s$, $B^*_s$, $D_s$, and $D^*_s$
mesons by the same procedure.
Summary and conclusions are given in Sec.~\ref{seconcl}.

%%%%%%%%%%%%%%%%%%%%%%%%%%%%%%
\section{Quark-meson coupling model}
\label{secqmc}
%%%%%%%%%%%%%%%%%%%%%%%%%%%%%%

The in-medium properties (Lorentz-scalar and Lorentz-vector potentials) of
the $B$, $B^*$, $D$, $D^*$, $K$ and $K^*$ mesons are calculated by the QMC model.
The original version of the QMC model describes the structure of
nucleons using the MIT bag model~\cite{Guichon:1987jp}, and the binding of the
nucleons by the scalar-isoscalar-$\sigma$, vector-isoscalar-$\omega$ and
vector-isovector-$\rho$ meson fields directly coupled to the
relativistically moving light $u$ and $d$ quarks
confined in the nucleon bags~\cite{Guichon:1987jp,Guichon:1995ue,Saito:1996sf}.
The mean fields are generated by the confined light quarks in the nucleons, and since the
(light-quark)-(meson field) couplings in any hadron should be the same as that in nucleon,
one can naturally expect that the hadrons with light quarks immersed in nuclear medium
to change their properties in the QMC model,
such as the internal structure and (Lorentz scalar effective) masses.

Let us consider the rest frame of nuclear matter, namely, the spin and isospin
saturated, uniformly distributed infinitely large many-nucleon system.
The Dirac equations for the quarks and antiquarks in nuclear matter 
(no Coulomb force) are given in the QMC model assuming SU(2)
symmetry for the light quarks ($q = u$ or $d$ and
$m_q = m_u = m_d$)~\cite{Tsushima:2002cc,Tsushima:1997df}
%%%
\begin{eqnarray}
&&\left[i\gamma \cdot \partial_{x} - \left(m_{q} - V^{q}_{\sigma}\right)
\mp \gamma^{0} \left(V^{q}_{\omega} + \frac{1}{2}V^{q}_{\rho}\right)\right]
\begin{pmatrix}
        \psi_{u}\left(x\right)\\
        \psi_{\overline{u}}\left(x\right)
       \end{pmatrix} = 0,\\
&&\left[i\gamma \cdot \partial_{x} - \left(m_{q} - V^{q}_{\sigma}\right)
\mp \gamma^{0} \left(V^{q}_{\omega} - \frac{1}{2}V^{q}_{\rho}\right)\right]
\begin{pmatrix}
        \psi_{d}\left(x\right)\\
        \psi_{\overline{d}}\left(x\right)
       \end{pmatrix} = 0,\\
&&\left[i\gamma \cdot \partial_{x} - m_{Q}\right]\psi_{Q, \overline{Q}}\left(x\right) = 0,
%&&\left[i\gamma \cdot \partial_{x} - m_{Q}\right]\psi_{\overline{Q}}\left(x\right)
\end{eqnarray}
%%%
where $Q$ and $\overline{Q}$,
respectively, stand for ($s,c$ or $b$) and ($\bar{s}, \bar{c}$ or $\bar{b}$),
and the mean-field potentials for the light quark ($q$) in nuclear matter are defined by
$V^{q}_{\sigma} \equiv g^{q}_{\sigma}\sigma$, 
$V^{q}_{\omega} \equiv g^{q}_{\omega}\omega = g^q_\omega\, \delta^{\mu,0} \omega^\mu$, 
$V^{q}_{\rho} \equiv g^{q}_{\rho}b = g^q_\rho\, \delta^{i,3} \delta^{\mu,0} \rho^{i,\mu}$, 
with the $g^{q}_{\sigma}$, $g^{q}_{\omega}$ and
$g^{q}_{\rho}$ being the corresponding quark-meson coupling constants.

The eigenenergies for the quarks and antiquarks in a hadron
$h (= B, B^*, D, D^*, K, \rm or\, K^*)$ in units of 
the in-medium bag radius of hadron $h$, $1/R^{*}_{h}$~\footnote{We indicate the in-medium quantity
with an asterisk *, except for indicating the vector mesons, $B^*, D^*, K^*, B_c^*, B_s^*$, and
$D_s^*$, which may be clear.}
are
%%%%%%%%%%%%%%%%%%%%%%%
\begin{eqnarray}
&&\begin{pmatrix}
        \epsilon_{u}\\
        \epsilon_{\overline{u}}
       \end{pmatrix} = \Omega^{*}_{q} \pm R^{*}_{h} 
\left(V^{q}_{\omega} + \frac{1}{2}V^{q}_{\rho}\right),\\
&&\begin{pmatrix}
        \epsilon_{d}\\
        \epsilon_{\overline{d}}
       \end{pmatrix} = \Omega^{*}_{q} \pm R^{*}_{h} 
\left(V^{q}_{\omega} - \frac{1}{2}V^{q}_{\rho}\right),\\
&&\epsilon_{s,c,b} = \epsilon_{\overline{s},\overline{c},\overline{b}} =
\Omega^*_{s,c,b}.
\end{eqnarray}
%%%%%%%%%%%%%%%%%%%%%%

The in-medium mass of the hadron $h$, $m^*_h$, is calculated by
%%%%%%%%%%%%%%%%
\begin{equation}
m^{*}_{h} = \sum_{j=q,\overline{q},Q,\overline{Q}} 
\frac{n_{j}\Omega^{*}_{j}- Z_{h}}{R^{*}_{h}} + \frac{4}{3} \pi R^{*3}_{h}B_{p},
\qquad
\left. \frac{d m^{*}_{h}}{d R_{h}}\right|_{R_{h} = R^{*}_{h}} = 0,
\end{equation}
%%%%%%%%%%%%%%%%
with $\Omega^{*}_{q} = \Omega^{*}_{\overline{q}} = \left[x^{2}_{q}
+ \left(R^{*}_h m^{*}_{q}\right)^{2}\right]^{\frac{1}{2}}$,
where $m^{*}_{q} = m_{q} - g^{q}_{\sigma}\sigma$
and $\Omega^{*}_{Q} = \Omega^{*}_{\overline{Q}} = \left[x^{2}_{Q} + \left(R^{*}_{h} 
m_{Q}\right)^{2}\right]^{\frac{1}{2}}$, with $x_{q,Q}$ being the lowest mode bag eigenfrequencies. 
Note that, due to the in-medium modification of $R^*_h$, $x_Q$ is also modified
slightly from the free space value, and thus, $\Omega^*_Q$ of the heavier quark
is also modified from that in free space.
$B_{p}$ is the bag constant, and $n_{q,Q}$ ($n_{\overline{q},\overline{Q}}$) are the lowest 
mode quark (antiquark) numbers for the quark flavors $q$ and $Q$ in the hadron $h$,
and the $Z_{h}$ parametrizes the sum of the center-of-mass
and gluon fluctuation effects and is
assumed to be independent of density~\cite{Guichon:1995ue}.

The current quark mass values used are
$(m_q, m_s, m_c, m_b) = (5, 250, 1270, 4200)$ MeV.
(See Ref.~\cite{Tsushima:2020gun} for the other values used,
$(m_q, m_s, m_c, m_b) = (5, 93, 1270, 4180)$ MeV.)
The free space nucleon bag radius is chosen to be
$R_{N}$ = 0.8 fm, and the light quark-meson coupling constants,
$g^{q}_{\sigma}$, $g^{q}_{\omega}$ and $g^{q}_{\rho}$, are determined by
the fit to the saturation energy (-15.7 MeV)
at the saturation density ($\rho_0 = 0.15$ fm$^{-3}$) of
symmetric nuclear matter, and the bulk symmetry energy 
(35 MeV)~\cite{Guichon:1987jp,Saito:2005rv}.

Note that the two-heavy flavored meson interactions with the nuclear matter
come from the vector and scalar interactions
at the hadronic level via the light-quark interaction
with the nuclear matter through the intermediate state heavy-light mesons,
where the light quark couplings with those fields as well as
in the nucleon
are determined to reproduce the symmetric nuclear matter saturation properties
using the Fermi gas description (with Fermi momenta of nucleons)~\cite{Guichon:1987jp}.
Thus, the repulsive vector interaction case does not correspond
directly to the quark Pauli blocking effect, and the present study is concerned with nuclear matter, not quark matter.

In Table~\ref{mesonmass}, we summarize the free space meson mass values (input)
taken from Particle Data Group (PDG)~\cite{ParticleDataGroup:2022pth}
except for the $m_{B^*_c}$ value, where we use the average value for
$m_{B^*_c}$ from Table III in Ref.~\cite{Martin-Gonzalez:2022qwd},
as well as the in-medium mass values for some mesons for $\rho_0, 2\rho_0$,
and $3\rho_0$ ($\rho_0=0.15$ fm$^{-3}$) calculated by the QMC model.
%%%%%%%%%%%%%%%%%%%%%%%%%%%%%%%%%%%%%%%%%%%%%%%%%%%%%%%%%%%%%%%%%%%%%%%
\begin{table}[htb!]
\begin{center}
\caption{Meson mass values (MeV) in free space (input) and at densities $\rho_0, 2\rho_0$ and
$3\rho_0$ ($\rho_0 = 0.15$fm$^{-3}$) calculated by the QMC model
with $(m_{u,d}, m_s, m_c, m_b) = (5, 250,1270,4200)$ MeV.
The free space values are from Particle Data Group (PDG)~\cite{ParticleDataGroup:2022pth}
except for $B^*_c$, where we use the average value for $m_{B^*_c}$ from Table III
in Ref.~\cite{Martin-Gonzalez:2022qwd}.
}
\label{mesonmass}
\vspace{1ex}
\begin{tabular}{c|c|c|c|c}
\hline
\hline
           &$\rho_B=0$ (input, MeV) &$\rho_B=\rho_0$ (MeV) &$\rho_B=2\rho_0$ (MeV)
&$\rho_B=3\rho_0$ (MeV)\\
\hline
\hline
%$m_\omega$  &782.7  &659.5  &590.7  &546.6  \\
%$m_\rho$    &775.3  &651.5  &582.3  &538.0  \\
$m_K$       &493.7  &430.5  &393.6  &369.0  \\
$m_{K^*}$   &893.9  &831.9  &797.2  &775.0  \\
%$m_\eta$    &547.9  &481.5  &443.0  &417.5  \\
%$m_{\eta'}$ &957.8  &899.3  &866.8  &846.0  \\
$m_D$       &1867.2 &1805.2 &1770.6 &1748.4 \\
$m_{D^*}$   &2008.6 &1946.9 &1912.9 &1891.2 \\
$m_B$       &5279.3 &5218.2 &5185.1 &5164.4 \\
$m_{B^*}$   &5324.7 &5263.7 &5230.7 &5210.2 \\
$m_{B_{c}}$   &6274.5  &  &  &  \\
$m_{B^{*}_{c}}$   &6333.0  &  &  &  \\
$m_{B^0_{s}}$   &5366.9  &  &  &  \\
$m_{B^{*}_{s}}$   &5415.4  &  &  &  \\
$m_{D_{s}}$ &1968.4  &  &  &  \\
$m_{D^{*\pm}_{s}}$ &2112.2  &  &  &  \\
\hline
\hline
\end{tabular}
\end{center}
\end{table}
%
%%%%%%%%%%%%%%%%%%%%%%%%%%%%%%%%%%%%%%%%%%%%%%%%%%%%%%%%%%%%%%%%%%%%%%%

\begin{figure}[htb!]
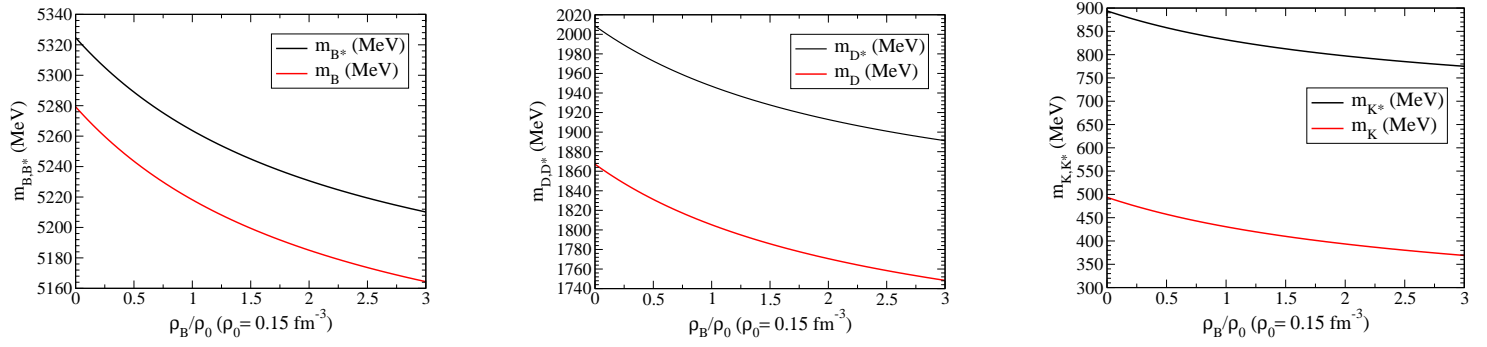
%
\vspace{4ex}
%\centering
\hspace{-16ex}
\includegraphics[width=5.6cm]{meson_BBs_mass.eps}
\hspace{8ex}
\includegraphics[width=5.6cm]{meson_DDs_mass.eps}
\hspace{8ex}
\includegraphics[width=5.6cm]{meson_KKs_mass.eps}
%\\
%\vspace{8ex}
%\hspace{-16ex}
%\includegraphics[width=8.0cm]{meson_KKs_mass.eps}
\caption{$B$ and $B^*$ (left panel), $D$ and $D^*$ (middle panel), and $K$ and $K^*$
(right panel) meson Lorentz-scalar effective masses in symmetric nuclear matter
versus baryon density ($\rho_B/\rho_0$), calculated by the QMC model.}%
\label{bksmass}
\end{figure}

In Fig.~\ref{bksmass}, we present the density dependence of the in-medium Lorentz-scalar effective
masses of the mesons, $B$, $B^*$, $D$, $D^*$, $K$, and $K^*$.
(Since the vector potentials will be canceled out for estimating the self-energy
of two-flavored heavy mesons in symmetric nuclear matter, we do not show them.)
The QMC model predicts similar mass shift of these mesons in symmetric nuclear matter
at $\rho_0$, and $\Delta m \equiv m^* - m$ are, respectively,
$(\Delta m_B, \Delta m_{B^*}, \Delta m_D, \Delta m_{D^*}, \Delta m_K, \Delta m_{K^*}) =
(-61.13, -61.05, -61.97, -61.66, -63.20, -61.97)$ MeV.
We use the density dependent in-medium masses of these mesons to estimate
the $B_c$, $B^*_c$, $B_s$, $B^*_s$, $D_s$,
and $D^*_s$ self-energies in symmetric nuclear matter.

Because the mass shift of the heavy-light meson will play a crucial role for the estimates of the
two-flavored heavy mesons in the present study, we would like to comment
that the mass shift of the heavy-light mesons might be controversial, as the
results of some studies do not show the downward mass shift of such mesons, especially the 
$D$ meson~\cite{Tolos:2004yg} and~\cite{Blaschke:2011yv},
where other newer studies predict the downward mass shift of the
$D$ meson including one of the same
author Tolos~\cite{Tolos:2009nn,Garcia-Recio:2011jcj,Albaladejo:2021cxj,
Mishra:2003se,Hayashigaki:2000es,Kumar:2019axp,Hilger:2008jg,Azizi:2014bba},
as well as the $B$ meson~\cite{Hilger:2008jg,Azizi:2014bba}.
(About the $D^*$ meson, one of them predicts a positive mass shift~\cite{Tolos:2009nn}.)
Although some of these studies predict negative mass shifts, those are less than
the ones obtained by the QMC model.
It is possible that, by including the quark Pauli blocking effect, the mass shift
presented in this study could be diminished. However, this is beyond our present approach.

Keeping these points in mind, we follow the present QMC model approach, and estimate the 
mass shift of the two-flavored heavy mesons in symmetric nuclear matter, 
encouraged by the major successful predictive features of the model
proven by experiments, $\Xi$ hypernuclei~\cite{Yoshimoto:2021ljs},
and in-medium modification of the bound proton electromagnetic form 
factors~\cite{Strauch:2003jx,JeffersonLabE93-049:2002asn,Dieterich:2000mu}.

%%%%%%%%%%%%%%%%%%%%%%%%%%%%%%%%%%%%%%%%%%%%
%%%%%%%%%%%%%%%%%%%%%%%%%%%%%%%%%%%%%%%%%

%%%%%%%%%%%%%%%%%%%%%%%%%%%%%%
\section{mass shift}
\label{secmshft}
%%%%%%%%%%%%%%%%%%%%%%%%%%%%%%

\begin{figure}[htb!]
%\begin{center}
\includegraphics[scale=0.25]{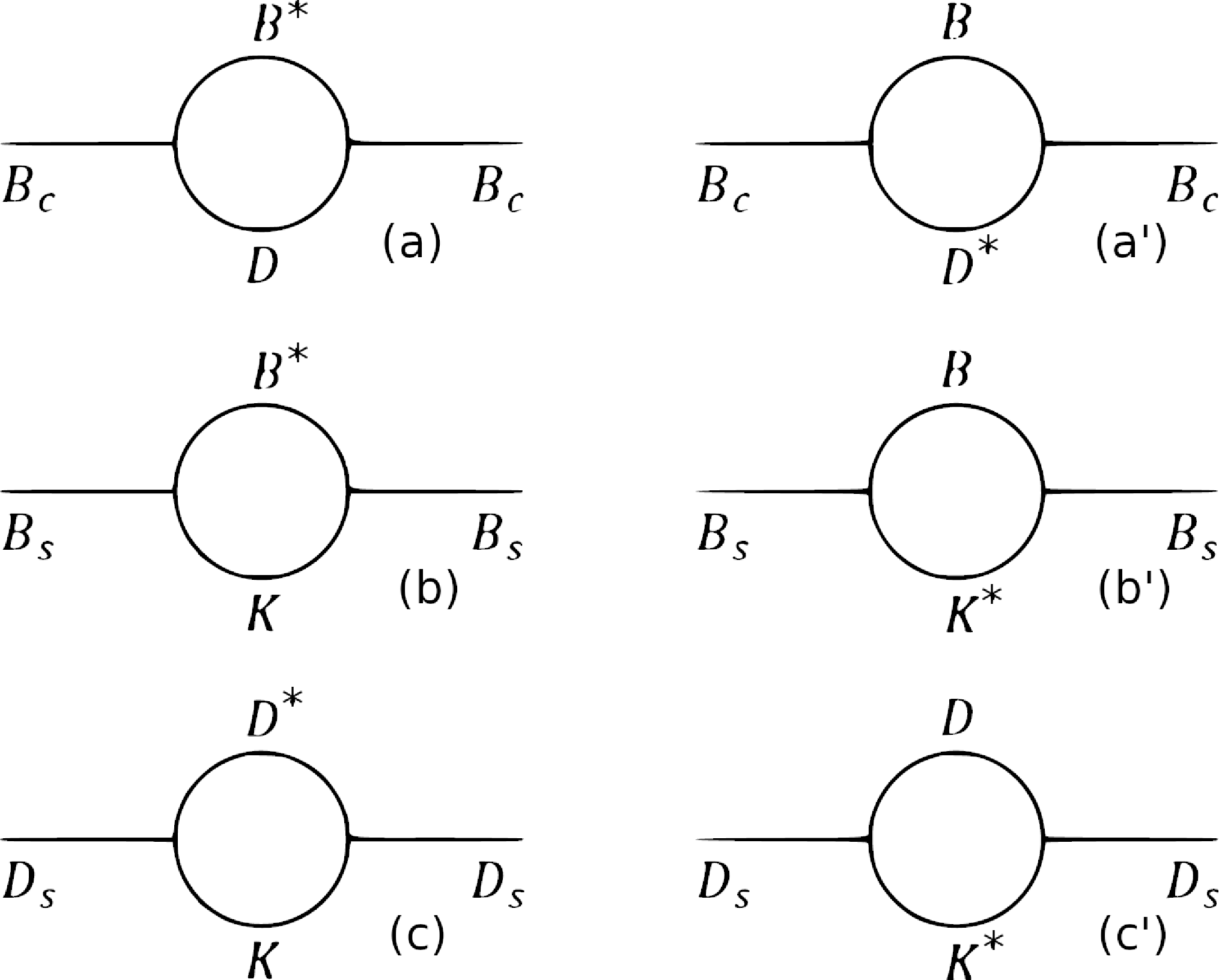}
\caption{Pseudoscalar meson self-energies for the mesons $B_c$ (top),
$B_s$ (middle), and $D_s$ (bottom), included in this study.
\label{psmeson}}
%\end{center}
\end{figure}

\begin{figure}[htb!]
\begin{center}
\includegraphics[scale=0.32]{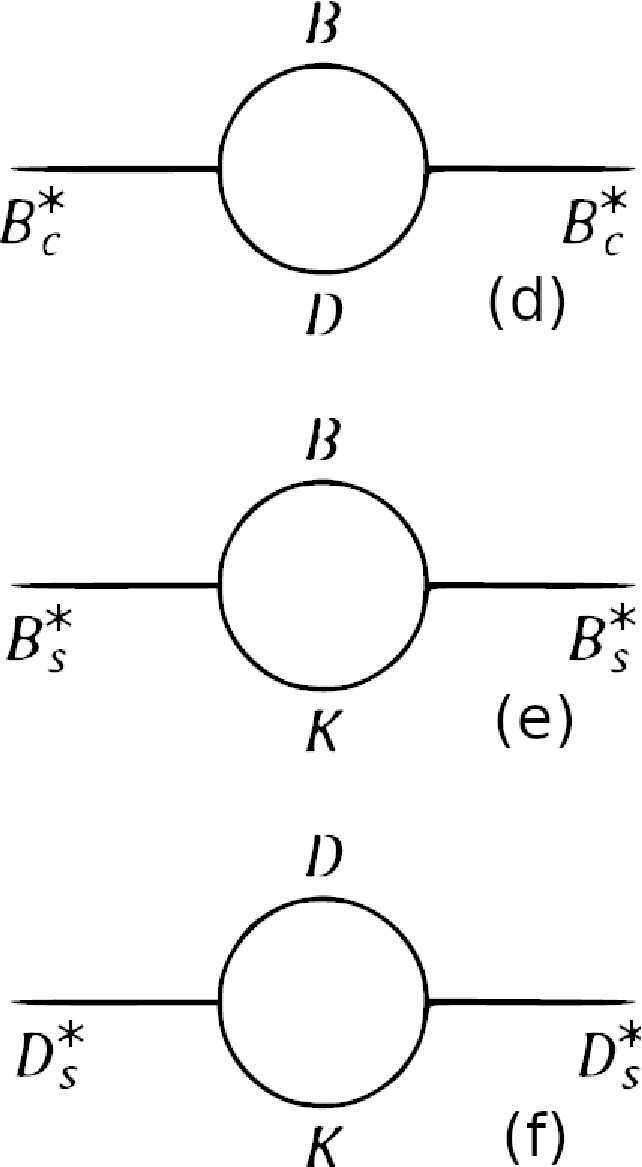}
\caption{Vector meson self-energies for the mesons
$B_{c}^{*}$ (top), $B_{s}^{*}$ (middle),
and $D_{s}^{*}$ (bottom) included in this study.
For the discussions on the self-energy, see text and Ref.~\cite{Zeminiani:2020aho}.
\label{vmeson}}
\end{center}
\end{figure}

In Figs.~\ref{psmeson} and~\ref{vmeson}, we, respectively show the self-energy graphs
considered in this study for pseudoscalar mesons $B_c, B_s$, and $D_s$, as well as
vector mesons $B_c^*, B_s^*$, and $D_s^*$.
For the vector meson self-energies, we include the ``lowest order graphs''
based on the discussions in Ref.~\cite{Zeminiani:2020aho},
namely, we exclude possible intermediate
two vector meson excitation graphs in each meson
because the heavier $B^*B^*$ (vector-vector)
loop contribution in the $\Upsilon$ self-energy (vector-vector-vector coupling)
gave unexpectedly larger contribution than those of the lighter $BB$ and $BB^*$ loop contributions.
(Probably we need more proper regularization methods and/or form factors.)
Thus, we intended to give the ``minimal'' in-medium mass
shift in Ref.~\cite{Zeminiani:2020aho}.
We follow this practice in the present study for consistency and also to allow us to compare the in-medium mass shift of the present study
directly with those of the quarkonia.
(See Ref.~\cite{Zeminiani:2020aho} for the detailed discussions on the issue.)
%%%%%%%%%%%%%%%%%%%%%%%%%%%%%%%%%%%%%%%%%%%%%%%%%%%%%%%%%%%%%%%
%%%%%%%%%%%%%%%%%%%%%%%%%%%%%%%%%%%%%%%%%%%%%%%%%%%%%%%%%%%%%%%

%%%%%%%%%%%%%%%%%%%%%%%%%%%%%%%%%%%%%%%%%%%%%%%%%%%%%%%%%%%%%%%
%%%%%%%%%%%%%%%%%%%%%%%%%%%%%%%%%%%%%%%%%%%%%%%%%%%%%%%%%%%%%

Concerning the pseudoscalar meson self-energy graphs in Fig.~\ref{psmeson},
it would be convenient to give the total mass (sum of the masses) of
the intermediate state mesons.
By this, one may naively guess about which graph will give dominant contribution
for each meson's self-energy,
based on a simple (nonrelativistic) quantum mechanical perturbation picture.
The total masses of the intermediate states in Fig.~\ref{psmeson} are

%%%%%%%%%%
\bea
(B^{*}D) \rightarrow m_{B^{*}} + m_D = 7191.9 {\rm\,\, MeV},
&&(BD^{*})\,\, {\rightarrow}\,\, m_B + m_{D^{*}} = 7287.9 {\rm\,\, MeV},
\label{tmass1}
\\
(B^{*}K) \rightarrow m_{B^{*}} + m_K = 5818.4 {\rm\,\, MeV},
&&(BK^{*})\,\, {\rightarrow}\,\, m_B + m_{K^{*}} = 6173.3 {\rm\,\, MeV},
\label{tmas2}
\\
(D^{*}K) \rightarrow m_{D^{*}} + m_K = 2502.3 {\rm\,\, MeV},
&&(DK^{*})\,\, {\rightarrow}\,\, m_D + m_{K^{*}} = 2761.1 {\rm\,\, MeV}.
\label{tmass3}
\eea
%%%%%%%%%%
One may expect that the intermediate states with lighter total mass state would give a dominant
contribution for the corresponding self-energy.
Note that in Eq.~(\ref{tmass1})
for the $(B^*D)$ and $(BD^*)$ loops, the total masses
are relatively close, and some other mechanisms
may influence the relative contributions easily.

However, it will turn out that the relativistic Lorentz structure of
the intermediate meson propagators, together with their masses,
will impact more than the total mass values for the self-energy.

The self-energies are calculated based on flavor SU(5) symmetric effective Lagrangian
densities~\cite{Lin:2000ke,Lodhi:2007zz} (hereafter called simply as Lagrangians).
The free Lagrangian for pseudoscalar and vector mesons
is given by~\cite{Lin:2000ke}
%%%%%%%%%%%%%%%%
\begin{equation}
{\cal L}_{0}=Tr \left(  \partial _{ \mu }P^{\dagger} \partial ^{ \mu }P \right)
-\frac{1}{2}Tr \left( F_{ \mu  \nu }^{\dagger}F^{ \mu  \nu } \right),
\end{equation}
%%%%%%%%%%%%%%
with \[ F_{ \mu  \nu }= \partial _{ \mu }V_{ \nu }- \partial _{ \nu }V_{ \mu } ,\]
where $P$ and $V$ (suppressing the Lorentz indices for $V$) are, respectively, the $5 \times 5$ pseudoscalar and vector meson matrices in SU(5):
%%%%%
\begin{eqnarray}
&&\hspace{-4ex} P = \frac{1}{\sqrt{2}} \begin{pmatrix} %%%\nonumber
\frac{\pi^{0}}{\sqrt{2}} + \frac{\eta}{\sqrt{6}} + \frac{\eta_{c}}{\sqrt{12}} 
+ \frac{\eta_{b}}{\sqrt{20}}  &  \pi^{+}  &  K^{+}  &  \overline{D}^{0}  &  B^{+}\\
\pi^{-}  &  \frac{-\pi^{0}}{\sqrt{2}} + \frac{\eta}{\sqrt{6}} + \frac{\eta_{c}}{\sqrt{12}} 
+ \frac{\eta_{b}}{\sqrt{20}}  &  K^{0}  &  D^{-}  &  B^{0}\\
K^{-}  &  \overline{K}^{0}  &  \frac{-2\eta}{\sqrt{6}} + \frac{\eta_{c}}{\sqrt{12}} 
+ \frac{\eta_{b}}{\sqrt{20}}  &  D_{s}^{-}  &  B_{s}^{0}\\
D^{0}  &  D^{+}  &  D_{s}^{+}  &  \frac{-3\eta_{c}}{\sqrt{12}} 
+ \frac{\eta_{b}}{\sqrt{20}}  &  B_{c}^{+}\\
B^{-}  &  \overline{B^{0}}  &  \overline{B_{s}^{0}}  &  B_{c}^{-}  &  \frac{-2\eta_{b}}{\sqrt{5}}\\
\end{pmatrix}, \nonumber \label{p} \\
\nonumber\\
\nonumber\\
&&\hspace{-4ex} V = \frac{1}{\sqrt{2}} \begin{pmatrix}
\frac{\rho^{0}}{\sqrt{2}} + \frac{\omega}{\sqrt{6}} + \frac{J/\Psi}{\sqrt{12}} 
+ \frac{\Upsilon}{\sqrt{20}}  &  \rho^{+}  &  K^{*+}  &  \overline{D}^{*0}  &  B^{*+}\\
\rho^{-}  &  \frac{-\rho^{0}}{\sqrt{2}} + \frac{\omega}{\sqrt{6}} + \frac{J/\Psi}{\sqrt{12}} 
+ \frac{\Upsilon}{\sqrt{20}}  &  K^{*0}  &  D^{*-}  &  B^{*0}\\
K^{*-}  &  \overline{K}^{*0}  &  \frac{-2\omega}{\sqrt{6}} + \frac{J/\Psi}{\sqrt{12}} 
+ \frac{\Upsilon}{\sqrt{20}}  &  D_{s}^{*-}  &  B_{s}^{*0}\\
D^{*0}  &  D^{*+}  &  D_{s}^{*+}  &  \frac{-3J/\Psi}{\sqrt{12}} + \frac{\Upsilon}{\sqrt{20}}  &  
B_{c}^{*+}\\
B^{*-}  &  \overline{B^{*0}}  &  \overline{B_{s}^{*0}}  &  B_{c}^{*-}  &  
\frac{-2\Upsilon}{\sqrt{5}} \\ 
\end{pmatrix}. \nonumber \label{v}
\end{eqnarray}

By the following minimal substitutions,
%%%%%
\begin{eqnarray}
&&\partial _{ \mu }P \rightarrow  \partial _{ \mu }P-\frac{ig}{2} \left[ V_{ \mu }\text{, P} 
\right],\\
&&F_{ \mu  \nu } \rightarrow  \partial _{ \mu }V_{ \nu }- \partial _{ \nu }V_{ \mu }
 -\frac{ig}{2} \left[ V_{ \mu },~V_{ \nu } \right],
\end{eqnarray}
%%%%%
one obtains the effective Lagrangian,
%%%%%
\begin{eqnarray}
\label{efflag}
 {\cal L}&={\cal L}_{0}+igTr \left(  \partial _{ \mu }P \left[ P,~V_{ \mu } \right]  \right) 
 -\frac{g^{2}}{4}Tr \left(  \left[ \text{P, V}_{ \mu } \right] ^{2} \right)\nonumber \\ 
 &+igTr \left(  \partial ^{ \mu }V^{ \nu } \left[ V_{ \mu },~V_{ \nu } \right]  \right) 
 +\frac{g^{2}}{8}Tr \left(  \left[ V_{ \mu },~V_{ \nu } \right] ^{2} \right),
\label{effLag}
\end{eqnarray}
%%%%%
where ${\cal L}_0$ is the free Lagrangian, and for the present study, we need
only the second term in Eq.~(\ref{effLag}), namely PPV
(pseudoscalar-pseudoscalar-vector) interactions.

%%%%%%%%%%%%%%%%%%%%%%%%%%%%%%%%%%%%%%%%%%%%%%%%%%%%%%%%%%%%%
\subsection{\boldmath{$B_c$} and \boldmath{$B^*_c$} meson results}
%%%%%%%%%%%%%%%%%%%%%%%%%%%%%%%%%%%%%%%%%%%%%%%%%%%%%%%%%%%%%

We first focus on $B_c$ and $B_c^*$ mesons.
The $B_c$ ($B^*_c$) in-medium negative mass shift comes from
the enhanced $B^*D + BD^*$ ($BD$) loop contribution to the self-energy, relative to those in free space.
See the Figs. (a) and (a') in Fig.~\ref{psmeson} [(d) in Fig.~\ref{vmeson}].
By expanding the Lagrangian Eq.~(\ref{efflag}) in terms of
the components of $P$ and $V$ matrices, we obtain the following
Lagrangians for the interactions $B_c B^*D$, $B_c BD^*$, and
$B^*_c BD$~\cite{Lodhi:2007zz}:
%%%%%%
\begin{eqnarray}
    \mathcal{L}_{B_{c}B^{*}D} &=& ig_{B_{c}B^{*}D}
    [(\partial_{\mu}B^{-}_{c}){D}
    - B^{-}_{c}(\partial_{\mu}{D})] B^{* \mu} + H.c.,
    \nonumber \\
    \mathcal{L}_{B_{c}BD^{*}} &=& ig_{B_{c}BD^{*}}
    [(\partial_{\mu}B^{+}_{c})\overline{B}
    - B_c^{+}(\partial_{\mu}\overline{B})] \overline{D^{*}}^\mu + H.c.,
   \nonumber \\
   \mathcal{L}_{B^{*}_{c}BD} &=& -ig_{B^{*}_{c}BD}
     B^{*+{\mu}}_{c} [\overline{B}(\partial_{\mu} \overline{D}) -
(\partial_{\mu} \overline{B}) \overline{D}] + H.c.,
\end{eqnarray}
%%%%%%
where the following conventions are used
%%%%%%
\begin{align*}
B&=\begin{pmatrix}
        B^{+}\\
        B^{0}
       \end{pmatrix}, & \overline{B}=\begin{pmatrix}
       B^{-} & \overline{B}^{0} \end{pmatrix},
			& &B^{*} =\begin{pmatrix}
        B^{*+}\\
        B^{*0}
       \end{pmatrix}, & &\overline{B^{*}}&=\begin{pmatrix}
       B^{*-} & \overline{B}^{*0} \end{pmatrix},\\
\end{align*}
%%%%%%

%%%%%%
\begin{align*}
\overline{D}&=\begin{pmatrix}
        \overline{D}^{0}\\
        D^{-}
       \end{pmatrix}, & D=\begin{pmatrix}
       D^{0} & D^{+} \end{pmatrix},  
			& &\overline{D}^{*} =\begin{pmatrix}
        \overline{D}^{*0}\\
        D^{*-}
       \end{pmatrix}, & &D^{*}&=\begin{pmatrix}
       D^{*0} & D^{*+} \end{pmatrix}.\\         
\end{align*}
%%%%%%

The SU(5) symmetric universal coupling $g$ in Eq.~(\ref{effLag})
yields the relations,
$g_{B_c B^* D} = g_{B_c B D^*} = g_{B^*_c B D}$.
The value of $g$ is fixed by
$g_{\Upsilon BB} = \frac{5g}{4\sqrt{10}} \approx 13.2$
by the $\Upsilon$ decay data $\Gamma(\Upsilon \to e^+ e^-)$
with the vector meson dominance (VMD) model~\cite{Lin:2000ke,Zeminiani:2020aho},
and thus, we get
%%%%%
\begin{equation}
g_{B_c B^* D} = \frac{2}{\sqrt{5}}g_{\Upsilon BB}, \hspace{3ex}
g_{B_c B^* D} = g_{B_c B D^*} = g_{B^*_c B D} = \frac{g}{2\sqrt{2}}
\approx 11.9.
\end{equation}
%%%%%

The in-medium Lorentz-scalar potential for the $B_c$ meson is computed by the difference of the
in-medium $m^*_{B_c}$ and the free space $m_{B_c}$ masses
%%%%%
\begin{equation}
V = m^*_{B_c} - m_{B_c},
\end{equation}
%%%%%%
where, the free space mass $m_{B_c}$ (input) is used to determine the bare mass $m^0_{B_c}$ by
%%%%%
\begin{equation}
m^2_{B_c} = \left( m^0_{B_c} \right)^2 - |\Sigma_{B_c}(k^2 = m^2_{B_c})|.
\label{m0}
\end{equation}
%%%%%%
Note that, the total self-energy $\Sigma_{B_c}$
is calculated by the sum of the $B^*D$ and $BD^*$ meson loop contributions in free space
ignoring the possible $B_c$ meson as well as all the other meson widths (or imaginary part)
in the self-energy.
The in-medium $B_c$ mass $m^{* 2}_{B_c}$
is similarly calculated, with the same bare mass
value $m^0_{B_c}$ determined in free space,
and the in-medium masses of the
($B,B^*,D,D^*$) mesons ($m^*_{B},m^*_{B^*},m^*_{D},m^*_{D^*}$), namely,
%%%%%
\bea
m^2_{B_c}
&=& \left[ m^0_{B_c}(B^*D+BD^*) \right]^2
- \left|\Sigma_{B_c}(B^*D) + \Sigma_{B_c}(BD^*) \right|(k^2 = m^2_{B_c}),
\label{m02}\\
m^{* 2}_{B_c}
&=& \left[ m^0_{B_c}(B^*D+BD^*) \right]^2
- \left|\Sigma^*_{B_c}(B^*D) + \Sigma^*_{B_c}(BD^*) \right|(k^{* 2} = m^{* 2}_{B_c}).
\label{m03}
\eea
%%%%%

We note that, when the self-energy graphs contain different contributions,
as $\Sigma_{B_c} ({\rm total}) = \Sigma(B^*D) + \Sigma(BD^*)$,
$m^0$ depends on both $\Sigma(B^*D)$ and $\Sigma(BD^*)$ to reproduce the physical mass
$m_{B_c}$. Thus, one must be careful when discussing the $B_c$
in-medium mass and mass shift of each loop contribution
$\Sigma(B^*D)$ and $\Sigma(BD^*)$ to be shown in Fig.~\ref{partbc}
since $m^0(B^*D+BD^*) \ne m^0(B^*D) \ne m^0(BD^*)$,
and $m^0(B^*D + BD^*) \ne m^0(B^*D) + m^0(BD^*)$.
The dominant loop contribution can be known by the decomposition
of the self-energy
$\Sigma^{(*)}_{B_c} (B^*D + BD^*) = \Sigma^{(*)}_{B_c}(B^*D) + \Sigma^{(*)}_{B_c}(BD^*)$,
which will be shown in Fig.~\ref{bcse}.

Considering solely the $B^*D$ loop without the $BD^*$ loop,
the $B_c$ self-energy is given by
%%%%%%
\begin{equation}
    \Sigma^{B^{*}D}_{B_{c}}(m^{*}_{B_{c}}) = \frac {-4g^{2}_{B_{c}B^{*}D}}{\pi^{2}} \int d|\textbf{k}| 
    |\textbf{k}|^{2} I_{B_c}^{B^{*}D}(|\textbf{k}|)
    F_{B_c B^* D} (\textbf{k}^2),
\label{SigBsD}
\end{equation}
where $I_{B_c}^{B^{*}D} (|\textbf{k}|)$ is expressed,
after the Cauchy integral with respect to $k^0$ complex plane shifting $k^0$ variable for the vector potential as,
%%%%%
\begin{eqnarray}
    I_{B_c}^{B^{*}D} (|\textbf{k}|) &=& 
    %\left(\frac{m^{2}_{B_{c}}}{m^{2}_{B^{*}}} \right) % \cdot 
    %\left[\frac{(-\omega^{*}_{B^{*}}+m^{*}_{B^{*}})
    %(-\omega^{*}_{B^{*}}-m^{*}_{B^{*}})}{(-2\omega^{*}_{B^{*}})
    %(-\omega^{*}_{B^{*}}-m^{*}_{B_{c}}+\omega^{*}_{D})
    %(-\omega^{*}_{B^{*}}-m^{*}_{B_{c}}-\omega^{*}_{D})}\right.
    %\nonumber \\
    %&+& \left. \frac{(m^{*}_{B_{c}}-\omega^{*}_{D}+m^{*}_{B^{*}})
    %(m^{*}_{B_{c}}
    %-\omega^{*}_{D}-m^{*}_{B^{*}})}{(m^{*}_{B_{c}}-\omega^{*}_{D}+
    %\omega^{*}_{B^{*}})(m^{*}_{B_{c}}-\omega^{*}_{D}-
    %\omega^{*}_{B^{*}})(-2\omega^{*}_{D})}\right]
    \left. \frac{m^{*2}_{B_c} \left(-1 + k^2_0 / m^{*2}_{B^*} \right)}
    {(k_0 - \omega^*_{B^*}) (k_0 - m^{*}_{B_c} + \omega^*_D)
    (k_0 - m^{*}_{B_c} -\omega^*_D)}
    \right|_{k_{0} = -\omega^*_{B^*} }
    \nonumber \\
    && \left. \hspace{5ex} + \frac{m^{*2}_{B_c} \left( -1 + k^2_0 / m^{*2}_{B^*} \right) }
    {(k_0 + \omega^*_{B^*}) (k_0 - \omega^*_{B^*})
    (k_0 - m^{*}_{B_c} -\omega^*_D)}
    \right|_{k_{0} = m^{*}_{B_c} - \omega^*_D},
\label{IBsD}
\end{eqnarray}
%%%%%%
where, $F_{B_c B^* D}$ in Eq.~(\ref{SigBsD}) is the product of vertex form factors
to regularize the divergence in the loop integral,
$F_{B_c B^* D} (\textbf{k}^2) = u_{B_cB^*}(\textbf{k}^2)u_{B_c D}(\textbf{k}^2)$.
They are given by
$u_{B_c B^*} = \left( \frac{\Lambda^2_{B^*} + m^2_{B_c}}
{\Lambda^2_{B^*} + 4\omega^2_{B^*} (\textbf{k}^2)} \right)^2$ and
$u_{B_c D} = \left( \frac{\Lambda^2_{D} + m^2_{B_c}}
{\Lambda^2_{D} + 4\omega^2_{D} (\textbf{k}^2)} \right)^2$
with $\Lambda_{B^*}$ and $\Lambda_{D}$ being the cutoff masses associated with the $B^*$ and $D$
mesons, respectively.
We use the common value $\Lambda = \Lambda_{B^*} = \Lambda_{D}$. 
A similar calculation is performed to obtain the $BD^*$ loop contribution,
namely, in Eqs.~(\ref{SigBsD}) and~(\ref{IBsD}), as well as in the form factors,
by replacing $(B^*,D) \to (B,D^*)$.

The choice of cutoff value has nonegligible impacts on the results.
In this study, we use the common cutoff $\Lambda \equiv \Lambda_{B,B^*,D,D^*,K,K^*}$
by varying the $\Lambda$ value. The $\Lambda$ value may be associated
with the energies to probe the internal structure of the mesons.
In the previous study~\cite{Zeminiani:2020aho}, it was observed that when the values
of the cutoff becomes close to the masses of the mesons in calculating
the self-energies, a certain larger cutoff mass value range did
not make sense to serve as the form factors.
This is because the Compton wavelengths of the corresponding
cutoff mass values reach near and/or smaller than those of the meson sizes.
Therefore, we need to constrain the $\Lambda$ value in such
a way that the form factors reflect properly the finite size of the mesons.
Based on the heavy quark and heavy meson symmetry in QCD,
we use the same range of values for $\Lambda$ as it was practiced for
the quarkonia~\cite{Zeminiani:2020aho}.
Thus, we use the values, $\Lambda$ = 2000, 3000, 4000, 5000, and 6000 MeV.

%%%%%%%%%%%%%%%%%%%%%%% Bc &&&&&&&&&&&&&&&&&&&&&&&&&&&&&&&&&&
\subsubsection{\boldmath{$B_c$} meson results}
%%%%%%%%%%%%%%%%%%%%%%%%%%%%%

We now present the $B_c$ meson in-medium mass and the mass shift.
(See Table~\ref{mesonmass} for the free space mass values of the mesons.)

In Table~\ref{Bcm0}, we give the cutoff ($\Lambda$) value dependence of the
$m^0$ in Eq.~(\ref{m0}) for different cases in the
self-energy calculations of $B_c$ meson, namely, only the $B^*D$ loop, only the $BD^*$ loop, and $B^*D + BD^*$ loops.
%%%%%
%%%%%%%%%%%%%%%%%%%%%%%%%%%%%%%%%%%%%%%
\begin{table}[htb!]
\caption{
Cutoff ($\Lambda$) value dependence of $m^0$ in Eq.~(\ref{m0})
for only the $B^*D$ loop, only the $BD^*$ loop, and $B^{*}D + BD^{*}$ loops.
\label{Bcm0}
}
\begin{center}   %%%%%%%%%%%%%%%%%%%%
\begin{tabular}{c|c|c|c}
\hline
\hline
$\Lambda$ (MeV)  &$m^0(B^{*}D)$ (MeV) &$m^0(BD^{*})$ (MeV)& $m^0(B^{*}D + BD^{*})$ (MeV)\\
\hline
\hline
2000		& 7906.1 &  9925.4 & 11029.6 \\
3000      	& 8032.6 & 10313.9 & 11468.7 \\
4000     	& 8249.0 & 10913.0 & 12156.0 \\
5000      	& 8561.5 & 11732.3 & 13098.8 \\
6000      	& 8968.6 & 12766.6 & 14284.6 \\
%%%2000		& 7906.14 MeV &  9925.44 MeV &  11029.6  MeV\\
%%%3000      	& 8032.64 MeV & 10313.94 MeV & 11468.71 MeV\\
%%%4000     	& 8248.97 MeV & 10912.96 MeV & 12156.02 MeV\\
%%%5000      	& 8561.51 MeV & 11732.32 MeV & 13098.75 MeV\\
%%%6000      	& 8968.57 MeV & 12766.56 MeV & 14284.64 MeV\\
\hline
\hline
\end{tabular}
\end{center}   %%%%%%%%%%%%%%%%%%%%%%%%%
\end{table}
%%%%%%%%%%%%%%%%%%%%%%%%%%%%%%%%%%%%%

\begin{figure}[htb]
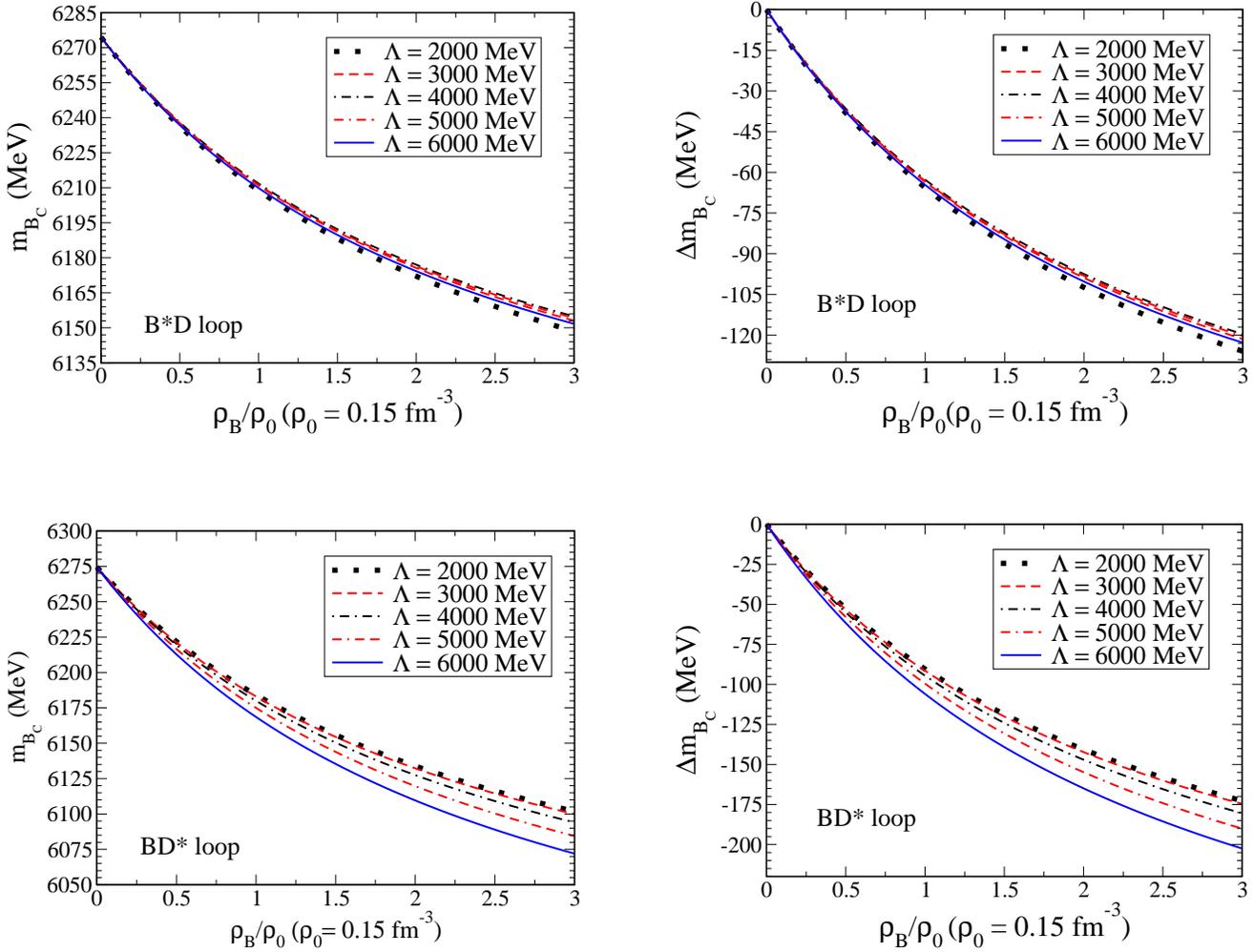
%
\vspace{8ex}
\hspace{-16ex}
\includegraphics[width=8.0cm]{Bc_BsD_mass.eps}
\hspace{8ex}
\includegraphics[width=8.0cm]{Bc_BsD_pot.eps}
\\
\vspace{8ex}
%\centering
\hspace{-16ex}
\includegraphics[width=8.0cm]{Bc_BDs_mass.eps}
\hspace{8ex}
\includegraphics[width=8.0cm]{Bc_BDs_pot.eps}
%%%%%
\caption{In-medium mass (left panel) and mass shift (right panel)
of $B_c$ meson (i) only the $B^*D$ loop (top)
and (ii) only the $BD^*$ loop (bottom panel)
versus baryon density ($\rho_B/\rho_0$) for five different
values of the cutoff mass $\Lambda$.}%
\label{partbc}%
\end{figure}

\begin{figure}[htb]%
\vspace{8ex}
%\centering
\hspace{-16ex}
\includegraphics[width=8.0cm]{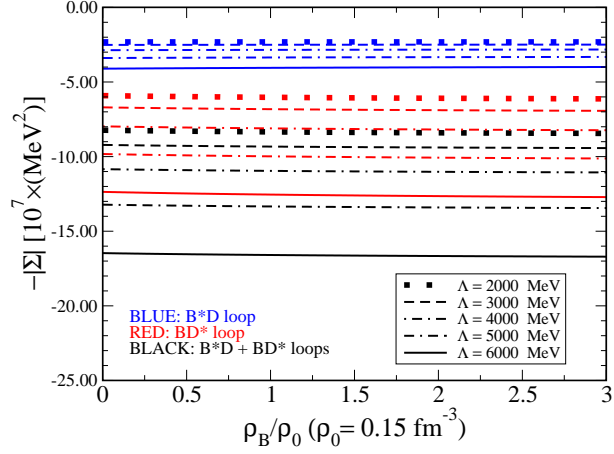}%
\caption{Decomposition of the $B_c$ meson total self-energy
$(B^*D + BD^*)$ loops
as a function of baryon density ($\rho_B/\rho_0$)
for five different values of the cutoff mass $\Lambda$.
}%
\label{bcse}%
\end{figure}

The in-medium mass and the mass shift of the $B_c$ meson
versus $\rho_B/\rho_0$ are presented in Fig.~\ref{partbc}
for (i) only the $B^*D$ loop (upper panel),
(ii) only the $BD^*$ loop (lower panel),
and (iii) for the $B^*D + BD^*$ loops, the decomposition
of $-|\Sigma|=|\Sigma_{B_c}(B^*D + BD^*)|$ in Fig.~\ref{bcse}, and
in-medium mass and the mass shift (our predictions) in Fig.~\ref{totbc}.
%%%%%
%%%%%%%%%%%%%%%%%%
%%%%%%%%%%%%%%%%%%%%%%%%%%%%%%%%%%%%%%%%%%%%%%%%%%%%%%%%%%%%%%%%%%

%%%%%%%%%%%%%%%%%%%%%%%%%%%%%
%%%%%%%%%%%%%%%%%%%%%

%%%%%%%%%%%%%%%%%%
\begin{figure}[htb]
\vspace{8ex}
%\centering
\hspace{-16ex}
 \includegraphics[width=8.0cm]{Bc_totalmass.eps}
\hspace{8ex}
 \includegraphics[width=8.0cm]{Bc_totalpot.eps}
 \caption{Total ($B^*D$ + $BD^*$) loop contribution for the
in-medium $B_c$ mass (left panel) and mass shift (right panel)
versus baryon density ($\rho_B/\rho_0$) for five different
values of the cutoff mass $\Lambda$.}
 \label{totbc}
\end{figure}
%%%%%%%%%%%%%%%%%

Comparing the results of $-|\Sigma|$ in Fig.~\ref{bcse},
we can conclude that the heavier total mass pair $BD^*$ loop contribution is dominant
over that of the lighter $B^*D$ pair loop.
This is different from our naive expectation based on the total mass values, that
the lighter total mass loop is expected to give the dominant contribution.
A possible explanation is in the following.
Looking at Eq.~(\ref{SigBsD}) for the $B^*D$ loop,
it contains the $B^*$ vector meson propagator, related to the expression
Eq.~(\ref{IBsD}). For the $BD^*$ loop, instead, one may replace $m^*_B \to m^*_D$ in
Eq.~(\ref{IBsD}) in the propagator.
Since $m_{B^*} - m_B \simeq$ 45 MeV $<< m_{B^*} - m_{D^*} \simeq 3300$ MeV
and $m_{D^*} - m_D \simeq 140$ MeV $<< m_{B^*} - m_{D^*} \simeq 3300$ MeV,
and, even in nuclear medium the $B^*$ and $D^*$ roles are changed,
the term $\propto k_0^2/m^{* 2}_{D^*}$ is expected to
give significant effect in the loop integral than that of
the term $\propto k_0^2/m^{* 2}_{B^*}$.
Thus, the conclusion is that the $D^*$ meson propagator,
the term $\propto k_0^2/m^{* 2}_{D^*}$, gives much larger contribution to the self-energy integral
than that of the $B^*$ propagator, $\propto k_0^2/m^{* 2}_{B^*}$.
We will also examine the other cases, namely, the self-energies of $B_s$ and $D_s$, whether or not
such things happen.
\\

\noindent $Summary$:\\
Including only the $BD^*$ loop for the $B_c$ self-energy with
the cutoff mass values $\Lambda$ = 2000 to 6000 MeV,
the $B_c$ mass shift $\Delta m_{B_c}(BD^*)$ at $\rho_0$ ranges from -90.4 to -106.0 MeV
[corresponding to the $B_c$ meson in-medium mass $m^*_{B_c}(BD^*)$ ranging from 6184.1 to 6168.5
MeV at $\rho_0$].
On the other hand, including only the $B^*D$ loop yields a less negative shift,
with $\Delta m_{B_c}(B^*D)$ ranging
from -65.7 to -64.6 MeV [$m^*_{B_c}(B^*D)$ ranging from
6208.8 to 6209.9 MeV] at $\rho_0$.
Interestingly, the larger cutoff value corresponds to a smaller mass shift.
This is because the cutoff value starts to get closer to the $B_c$ meson mass value,
which is around 6000 MeV, and the cutoff dependence gets smaller
until it reaches a point where the mass shift starts to increase
(become less negative).
This kind of behavior was already observed and discussed in Ref.~\cite{Zeminiani:2020aho},
when the cutoff $\Lambda$ value becomes too large.
It is expected that when the $\Lambda$ value becomes larger a certain value
(larger than 6000 MeV), the $B_c$ mass shift would turn to be positive (in fact it happens),
which is not proper, because the form factor does not make sense as already commented.
Finally, including the total ($BD^* + B^*D$) loop contributions (our prediction),
the $B_c$ mass shift $\Delta m_{B_c}(BD^*+B^*D)$
at $\rho_0$ ranges from -90.4 to -101.1 MeV
[$m^*_{B_c}(BD^* + B^*D) =$ 6184.1 to 6173.4 MeV].
Later, we will compare the $B_c$ mass shift and those of the $\eta_b$ and $\eta_c$.

%%%%%%%%%%%%%%%%%%%%%%% Bcs &&&&&&&&&&&&&&&&&&&&&&&&&&&&&&&&&&
\subsubsection{\boldmath{$B_c^*$} meson results}
%%%%%%%%%%%%%%%%%%%%%%%%%%%%%

Here, we present the in-medium mass and the mass shift of the $B^*_c$ meson.
For the $B^*_c$ self-energy, we include only the $BD$ loop contribution,
as already commented based on the $\Upsilon$ and $J/\psi$
self-energies~\cite{Zeminiani:2020aho},
%%%%%%%%%%%%%%%%%%%%%%%%
\begin{equation}
\Sigma^{BD}_{B^{*}_{c}}(m^*_{B_c^*}) = \frac{-4g^{2}_{B^{*}_{c}BD}}{3\pi^{2}} \int d|\textbf{k}|
|\textbf{k}| ^{4} I^{BD}_{B^{*}_{c}}(|\textbf{k}|)
F_{B^*_c BD} (\textbf{k}^2),
\label{IBcs}
\end{equation}
%%%%%%%%%%%%%%%%%%%%%%
where $I^{BD}_{B^{*}_{c}}(|\textbf{k}|)$ is expressed by
%%%%%%%%%%%%%%%%%%
\begin{eqnarray}
    I^{BD}_{B^{*}_{c}}(|\textbf{k}|) &=&
%%%\frac{1}{(-2 \omega^*_B) (-\omega^*_B - m^*_{B^*_c} +
%%%\omega^*_D) %(-\omega^*_B - m^*_{B^*_c} - %\omega^*_D)}
%%%\nonumber \\
%%%&+& \frac{1}{(m^*_{B^*_c} - \omega^*_D + \omega^*_B)
%%%(m^*_{B^*_c} - \omega^*_D - \omega^*_B) (-2 \omega^*_D)}
    \left. \frac{1}{(k_0 - \omega^*_B) (k_0 - m^*_{B^*_c} + \omega^*_D)
    (k_0 - m^*_{B^*_c} - \omega^*_D)} \right|_{k_0 = - \omega^*_B}
    \nonumber \\
    && \left. \hspace{5ex} + \frac{1}
    {(k_0 + \omega^*_B) (k_0 - \omega^*_B)
    (k_0 -m^*_{B^*_c} - \omega^*_D)} \right|_{k_0 = m^*_{B^*_c} - \omega^*_D}.
\end{eqnarray}
%%%%%%%%%%%%%%%%%%%%%%%%
In Eq.~(\ref{IBcs}), $F_{B^*_c BD} (\textbf{k}^2)$ is given by
the product of the form factors,
$F_{B^*_c BD} (\textbf{k}^2) = u_{B^*_c B}(\textbf{k}^2)
u_{B^*_c D}(\textbf{k}^2)$,
with $u_{B^*_c B}$ and $u_{B^*_c B}$ being
$u_{B^*_c B} = \left( \frac{\Lambda^2_{B} + m^2_{B^*_c}}
{\Lambda^2_{B} + 4\omega^2_{B} (\textbf{k}^2)} \right)^2$ and
$u_{B^*_c B} = \left( \frac{\Lambda^2_{D} + m^2_{B^*_c}}
{\Lambda^2_{D} + 4\omega^2_{D} (\textbf{k}^2)} \right)^2$.
Again we use $\Lambda = \Lambda_B = \Lambda_D$ ranging 2000 to 6000 MeV.

\begin{figure}[htb!]%
%\vspace{8ex}
%\centering
\hspace{-16ex}
\includegraphics[scale=0.4]{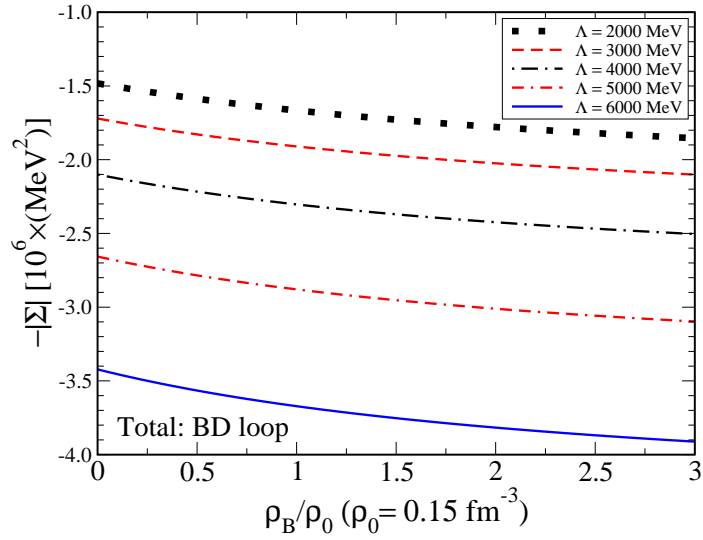}%
\caption{$BD$ loop (total) contribution for the $B^*_c$ meson self-energy as a function of baryon density ($\rho_B/\rho_0$) for five different values of the cutoff mass $\Lambda$.
}%
\vspace{5ex}
\label{bcsse}%
\end{figure}

\begin{figure}[htb!]
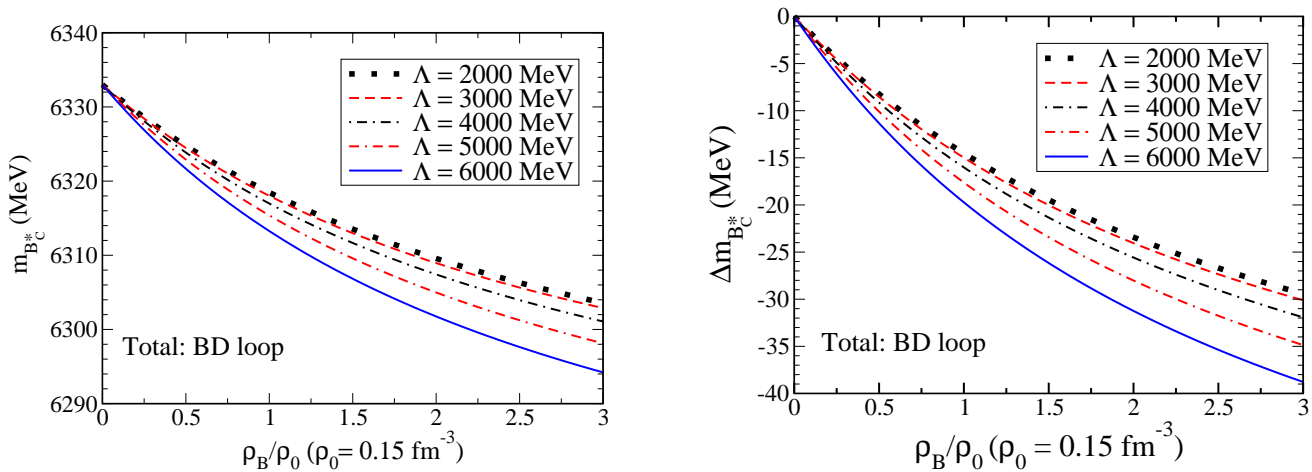

%\vspace{8ex}
%\centering
\hspace{-16ex}
 \includegraphics[width=8.0cm]{Bcs_mass.eps}
\hspace{8ex}
 \includegraphics[width=8.0cm]{Bcspot.eps}
 \caption{$BD$ loop (total) contribution for the
in-medium $B^*_c$ mass (left panel) and the mass shift (right panel) versus baryon density
($\rho_B/\rho_0$) for five different values of the cutoff mass $\Lambda$.}
 \label{totbcs}
\end{figure}

The cutoff mass value dependence of $m^0$ is given in Table~\ref{Bcsm0}, and
the results for the $BD$ (total) loop contribution for the $B^*_c$ self-energy
and mass shift in symmetric nuclear matter are shown in
Figs.~\ref{bcsse} and~\ref{totbcs}, respectively,
for five different values of $\Lambda$.
The mass shift $\Delta m_{Bc} (BD)$ at $\rho_0$ rages from -14.5 to -19.7 MeV,
corresponding to the in-medium masses $m^*_{Bc}(BD)$ from 6318.5 to 6313.3 MeV.
The mass shift is much less than that of the $B_c$ meson.
%%%%%%%%%%%%%%%%%%%%%%%%%%%%%%%%%%%%%%%%%%%%%%%%
\begin{table}[htb!]
\caption{
See caption of Table~\ref{Bcm0}, but for $B^*_c$ meson.
\label{Bcsm0}
}
\begin{center}   %%%%%%%%%%%%%%%%%%%%
\begin{tabular}{c|c}
\hline
\hline
$\Lambda$ (MeV)  	&    $m^0(BD)$ (MeV) \\
\hline
\hline
2000	& 6449.0 \\
3000    & 6467.4 \\
4000    & 6496.8 \\
5000    & 6539.4 \\
6000    & 6597.7 \\
%%%2000	& 6449.00 \\
%%%3000 & 6467.41 \\
%%%4000 & 6496.75 \\
%%%5000 & 6539.40 \\
%%%6000 & 6597.66 \\
\hline
\hline
\end{tabular}
\end{center}   %%%%%%%%%%%%%%%%%%%%%%%%%
\end{table}
%%%%%%%%%%%%%%%%%%%%%%%%%%%%%%%%%%%%%%%%%%%%%%%

%%%%%%%%%%%%%%%%%%%%%%%%%%%%%
%%%%%%%%%%%%%%%%%%%%%

%%%%%%%%%%%%%%%%%%
%%%%%%%%%%%%%%%%%

%%%%%%%%%%%%%%%%%%%%%%%%%%%%%%
\subsection{Comparison with heavy quarkonia}
\label{hvqrkcomp}
%%%%%%%%%%%%%%%%%%%%%%%%%%%%%%

\begin{figure}[htb!]
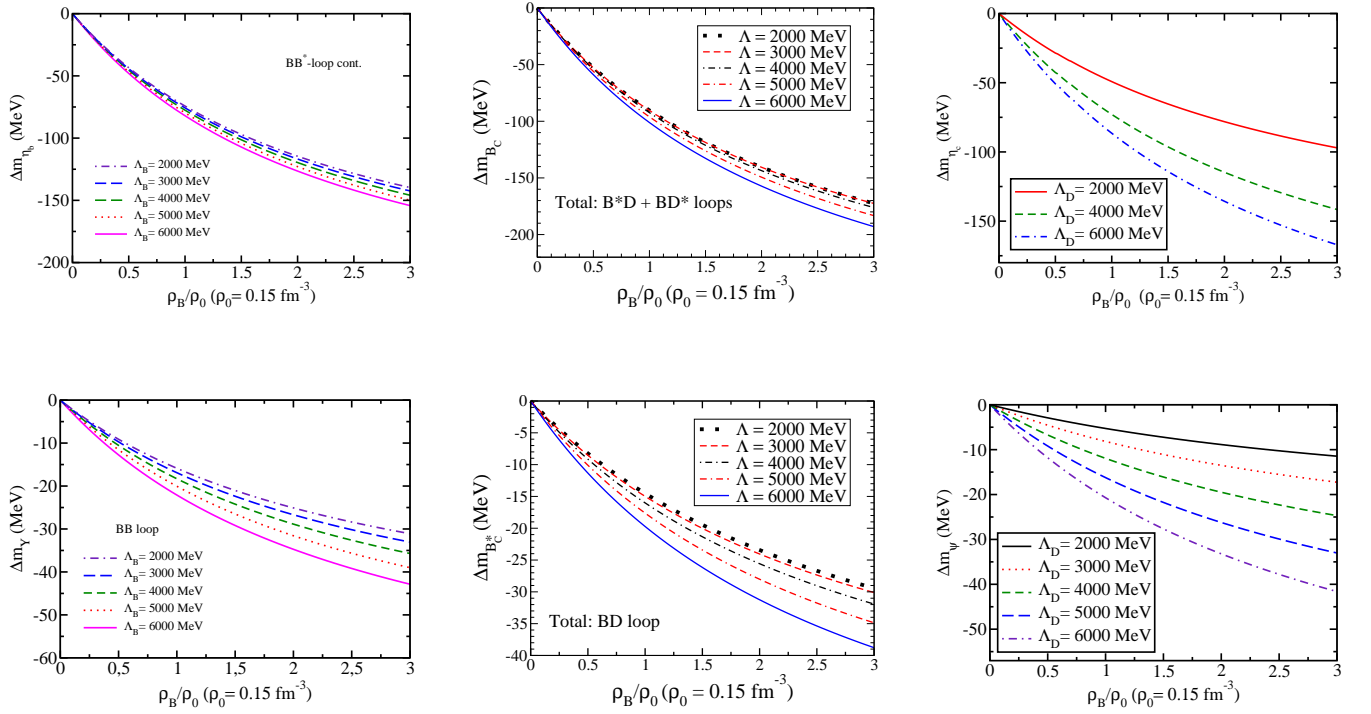
%
\vspace{8ex}
%\centering
%\hspace{-8ex}
\includegraphics[width=5.4cm]{etab_BBs.eps}
\hspace{4ex}
\includegraphics[width=5.4cm]{Bc_totalpot.eps}
\hspace{4ex}
\includegraphics[width=5.4cm]{Dmetac_DDsOnly.eps}
\\
\vspace{8ex}
%\hspace{-10ex}
\includegraphics[width=5.4cm]{1_BB.eps}
\hspace{4ex}
\includegraphics[width=5.4cm]{Bcspot.eps}
\hspace{4ex}
\includegraphics[width=5.4cm]{DJPsi_in_medium_XiOff_OnlyDDbar.eps}
\caption{Comparison of the mass shift of $B_c$ with $\eta_b$ and $\eta_c$ (upper panel)
as well as of $B^*_c$ with $\Upsilon$ and $J/\Psi$ (lower panel).}%
\label{comp}%
\end{figure}

We now compare in Fig.~\ref{comp} the results of $B_c$ and $B^*_c$
with those of the heavy
quarkonia~\cite{Krein:2010vp,Krein:2017usp,Zeminiani:2020aho,Cobos-Martinez:2020ynh}.
Since the $B_c$ meson is a pseudoscalar meson, we compare with the bottomonium $\eta_b$
and charmonium $\eta_c$ (upper panel), while for the $B_c^*$ meson, we compare with
those of the $\Upsilon$ and $J/\psi$ (lower panel).

For comparison, we would like to emphasize that the comparison
using the empirical extracted SU(4) sector coupling constants
for the charm sector would be more reasonable than using that empirically
extracted SU(5) sector coupling constant from the $\Gamma (\Upsilon \rightarrow e^+ e^-)$
since the SU(5) symmetry breaking is expected to be much larger than that of the SU(4)
symmetry based on the quark masses.

The value for the coupling constant of the vertex $J/\Psi DD$
used in the calculation of $J/\Psi$ mass shift was obtained from the
experimental data for $\Gamma (J/\Psi \rightarrow e^+ e^-)$ by the VMD hypothesis

\begin{equation}
\label{gJPsi}
    g_{J/\Psi DD} = \frac{g}{\sqrt{6}} \approx 7.7,
\end{equation}
where $g$ is the universal SU(4) coupling.

For the coupling constant $g_{\eta_{c} DD^{*}}$ used in the calculation of the $\eta_c$ mass shift,
we also adopt the SU(4) symmetry for the charm sector, which gives the relation

\begin{equation}
 g_{\eta_{c} DD^{*}} = g_{J/\Psi DD} = \frac{g}{\sqrt{6}} \approx 7.7.
\end{equation}

A comprehensive list of the values used for the coupling constants is presented in Table.~\ref{tblcpl45}.

%%%%%%%%%%%%%%%%%%%%%%%%%%%%%%%%%%%%%%%%%%%%%%%%%%%%%%%%%%%%%%%%%%%%%%%
\begin{table}
\caption{\label{tblcpl45} Coupling constant values in SU(4) and SU(5) symmetries.}
\begin{center}
%\scalebox{0.8}{
\begin{tabular}{ll|r}
  \hline \hline
  & & \multicolumn{1}{c}{SU(4)} \\
\hline
&$g$ & 18.9 \\
& $g_{J/\Psi DD}$ & 7.7 \\
& $g_{\eta_{c} DD^{*}}$ & 7.7 \\
  \hline \hline
  & & \multicolumn{1}{c}{SU(5)} \\
\hline
&$g$ & 33.4 \\
& $g_{\Upsilon BB}$ & 13.2 \\
& $g_{\eta_{b} BB^{*}}$ & 13.2 \\
& $g_{B_{c}B^{*}D}$ & 11.9 \\
& $g_{B^{*}_{s}BD}$ & 11.9 \\
\hline \hline
\end{tabular}
%}
\end{center}
\end{table}
%
%%%%%%%%%%%%%%%%%%%%%%%%%%%%%%%%%%%%%%%%%%%%%%%%%%%%%%%%%%%%%%%%%%%%%%%

Although we make this comparison, we admit
that this is not done based on a rigorous SU(5) symmetry of the same footing.
Repeatedly, the coupling constant $g$ is calculated
for the charm sector ($J/\Psi$, $\eta_c$) based on the SU(4) symmetry and
for the bottom sector ($\Upsilon$, $\eta_b$) and ($B_c$, $B^*_c$) based on the SU(5) symmetry.
This comparison would make a sense based on
the fact that SU(5) symmetry is much more broken
by the quark masses than that of SU(4).
Note that, although for the mass shift $\Delta m_{\eta_c}$~\cite{Cobos-Martinez:2020ynh},
the cutoff mass values $\Lambda =\Lambda_D=\Lambda_{D^*} = 3000$ and $5000$ MeV
are missing, it is irrelevant
to see the mass shift range for the cutoff range between the 2000 MeV and 6000 MeV.
%%%%%%%%%%%%%%%%%%
%%%%%%%%%%%%%%%%%%%%%%%%%%%%%%%%%%%%%%%%%%%%%%%%%%%%%%%%%%%%%%%%%%

In the study of the $\eta_c$ mass shift, only the $DD^*$ loop contribution was included,
and it corresponds to the mass shift
$\Delta m_{\eta_c} (DD^*)$
at $\rho_0$ ranges  -49.2 to -86.5,
for the cutoff mass values $\Lambda_D = \Lambda_{D^*}$ of 2000, 4000 and 6000 MeV.
The estimated values for the $\eta_b$ mass shift 
$\Delta m_{\eta_b} (BB^*)$
at $\rho_0$ including only the $BB^*$ loop, ranges from -74.2 to -82.0 MeV,
where the same range of the cutoff mass value is applied for the present study.
The $B^*D$ (heavier pair) contribution for the $B_c$
mass shift yields a less negative mass shift
compared with those of the $\eta_b$ and $\eta_c$ (except for $\Lambda$ = 2000 MeV).
On the other hand, only the $BD^*$ loop, as well as the total $B^*D + BD^*$ loop
contributions for the  $B_c$ mass shift give more negative mass shift
than those of the $\eta_b$ and $\eta_c$.
These facts indicate that the $B_c$ mass shift does not show the middle
range mass shift between those of the $\eta_c$ and $\eta_b$,
different from our naive expectation.

Next, we compare the mass shift of $\Upsilon, B_c^*$ and $J/\psi$ in Fig~\ref{comp}
(lower panel).
The $\Upsilon$ and $J/\Psi$ mass shift are obtained by taking, respectively, only the
(minimal) $BB$ and $DD$ loop contributions
corresponding to the present $B^*_c$ meson treatment with only the $BD$ loop.
The mass shift $\Delta m_{\Upsilon} (BB)$ at $\rho_0$ ranges from -15.9 to -22.1
MeV, while $\Delta m_{J/\psi} (DD)$ at $\rho_0$ ranges from
-5.3 to -20.7, when the common range
of the $\Lambda$ (2000 to 6000 MeV) is used.
The corresponding $B^*_c$ mass shift $\Delta m_{B_c^*} (BD)$
at $\rho_0$ ranges from -14.5 to - 19.7 MeV.
The $B^*_c$ meson in-medium mass shift is less dependent on the cutoff mass value than that of the
$J/\psi$.
Although the mass shift is dependent on the cutoff mass value,
the global trend shown in the lower panel of Fig.~\ref{comp}
indicates that the $\Delta m_{B^*_c}$ is more or less
the middle of the corresponding $\Delta m_\Upsilon$ and $\Delta m_{J/\Psi}$.
\\

\noindent
$Summary:$\\
Our results for the mass shift show the relations:
$|\Delta m_{\eta_c}| < |\Delta m_{\eta_b}| < |\Delta m_{B_c}|$
and $|\Delta m_{J/\Psi}| < |\Delta m_{B^*_c}| < |\Delta m_{\Upsilon}|$.

%%%%%%%%%%%%%%%%%%%%%%%%%%%%%%
\subsection{In-medium masses and mass shift of
\boldmath{$B_s$}, \boldmath{$B^*_s$}, \boldmath{$D_s$} and \boldmath{$D^*_s$}}
%%%%%%%%%%%%%%%%%%%%%%%%%%%%%%

In addition to the $B_c$ and $B^*_c$ mesons, we study here the in-medium mass
shift of the heavy-strange $B_s$ ($B^0_s$), $B^*_s$, $D_s$, and $D^*_s$ mesons.
The free space mass values of the $B_s$, $B^*_s$, $D_s$, and $D^*_s$ are
given in Table~\ref{mesonmass}.

We estimate the in-medium masses and mass shift of these mesons similarly
to those of the $B_c$ and $B^*_c$, where the interaction Lagrangians are obtained using the explicit components of the $P$ and $V$ matrices for the second term in Eq.~(\ref{effLag}), as before.
The relevant Lagrangians are
%%%%%%%%%%%%%%%%
\begin{eqnarray}
    \mathcal{L}_{B_sB^{*}K} &=& ig_{B_sB^{*}K}[(\partial_{\mu}\overline{B^0_s})\overline{K}
     - \overline{B^0_s} (\partial_{\mu}\overline{K})]{B}^{* \mu} + H.c.,
     \nonumber \\
     \mathcal{L}_{B_sBK^*} &=& ig_{B_sBK^*}[(\partial_{\mu}{B^0_s})\overline{B}
     - {B^0_s}(\partial_{\mu}\overline{B})]{K}^{* \mu} + H.c.,
     \nonumber \\
     \mathcal{L}_{B^{*}_{s}BK} &=& -ig_{B^{*}_{s}BK}
     {B_s^*}^{0 \mu} [\overline{B} (\partial_\mu K)
     - (\partial_\mu \overline{B}) K]  + H.c.,
     \nonumber \\
     \mathcal{L}_{D_sD^{*}K} &=& ig_{D_sD^{*}K}[(\partial_{\mu}D_{s}^+)\overline{K}
    - D_s^+ (\partial_\mu \overline{K})]\overline{D^*}^\mu + H.c.,
    \nonumber \\
    \mathcal{L}_{D_sDK^*} &=& ig_{D_sDK^*}[(\partial_{\mu}D^-_{s})D
    - D_{s}^- (\partial_{\mu}D)  ]{K}^{* {\mu}} + H.c.,
    \nonumber \\
    \mathcal{L}_{D^{*}_{s}DK} &=& -ig_{D^{*}_{s}DK}
     {D_s^{*-}}^\mu [D(\partial_{\mu}K)
     - (\partial_{\mu}D) K]  + H.c.
\end{eqnarray}
%%%%%%%%%%%%%%

The coupling constants for each vertex are obtained from the universal SU(5) coupling $g$ by
\begin{equation}
g_{B_{s} KB^*} = g_{B_{s} BK^*} = g_{B^{*}_{s} BK} =
g_{D_{s} KD^*} = g_{D_{s} DK^*} = g_{D^{*}_{s} DK} = 
\frac{g}{2\sqrt{2}} \approx 11.9.
\end{equation}

%%%%%%%%%%%%%%%%%%%%%%%%%%%%%%%%%%%%%%%%%%
\subsubsection{\boldmath{$B_s$} meson results}
%%%%%%%%%%%%%%%%%%%%%%%%%%%%%%%%%%%%%%%%%

We discuss first the $B_s$ meson.
In Table~\ref{Bs0m0}, we give the cutoff mass $\Lambda$ dependence
of the $m^0$ for each case of only the $B^*K$ loop, only the $BK^*$ and
$B^*K + BK^*$ loops. [See Eq.~(\ref{m0}).]
One can notice oscillating behavior of $m^0$
on $\Lambda$ value for all the cases.
Namely, first the $m^0$ value decreases from $\Lambda=2000$ MeV to
$\Lambda=3000$ MeV ($\Lambda=4000$ MeV for $B^*K$ loop) and increases
as $\Lambda$ value increases.
For the $B_c$ and $B_c^*$ cases, the behavior is
to monotonically increase as the $\Lambda$ value increases.
(Such behavior will also be noticed for the $B_s^*$ case with the $BK$ loop.)
%%%%%
%%%%%%%%%%%%%%%%%%%%%%%%%%%%%%%%%%%%%%%%%%%%%%%%%%%%%%%%%%%%%%%%%%%%%%%%%
\begin{table}[htb!]
\caption{
Cutoff $\Lambda$ value dependence of $m^0$ for only the $B^*K$ loop,
only the $BK^*$ loop, and total $B^{*}K + BK^{*}$ loops.
\label{Bs0m0}
}
\begin{center}   %%%%%%%%%%%%%%%%%%%%
\begin{tabular}{c|c|c|c}
\hline
\hline
$\Lambda$ (MeV) &$m^0(B^{*}K)$ (MeV) &$m^0(BK^{*})$ (MeV) &$m^0(B^{*}K + BK^{*})$ (MeV)\\
\hline
\hline
2000		& 8114.6 & 14394.5 & 15628.3 \\
3000      	& 7536.4 & 14358.3 & 15302.1 \\
4000     	& 7441.7 & 15180.7 & 16032.1 \\
5000      	& 7592.3 & 16688.6 & 17531.4 \\
6000      	& 7899.3 & 18750.5 & 19625.9 \\
\hline
\hline
\end{tabular}
\end{center}   %%%%%%%%%%%%%%%%%%%%%%%%%
\end{table}
%%%%%%%%%%%%%%%%%%%%%%%%%%%%%%%%%%%%%%%%%%%%%%%%%%%%%%%%%%%%%%%%%%%%%%%%

\begin{figure}[htb]
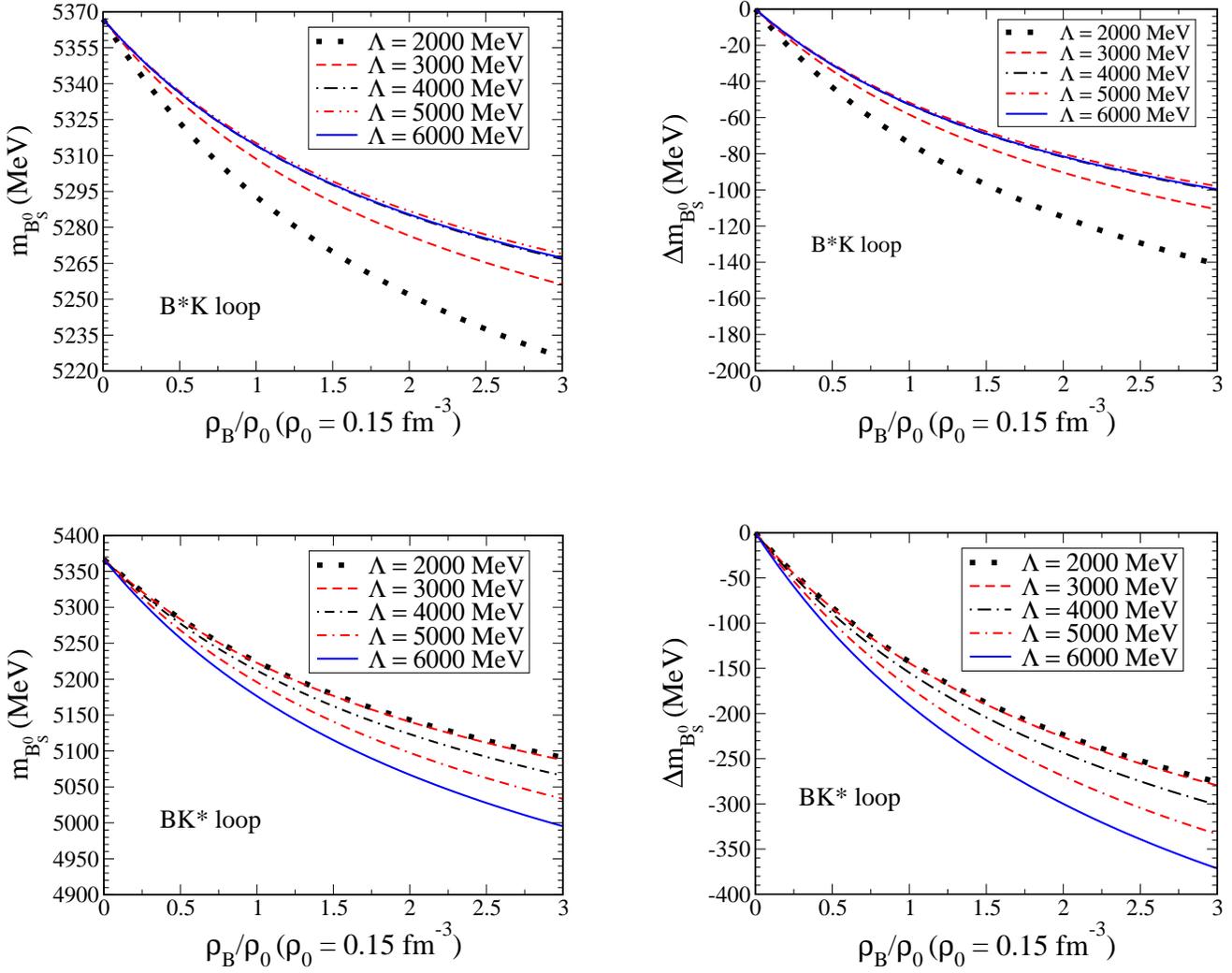
%
\vspace{8ex}
%\centering
\hspace{-16ex}
\includegraphics[width=8.0cm]{Bs0_BsK_mass.eps}
\hspace{8ex}
\includegraphics[width=8.0cm]{Bs0_BsK_pot.eps}
\\
\vspace{8ex}
\hspace{-16ex}
\includegraphics[width=8.0cm]{Bs0_BKs_mass.eps}
\hspace{8ex}
\includegraphics[width=8.0cm]{Bs0_BKs_pot.eps}
\caption{In-medium mass (left panel) and the mass shift (right panel)
of the $B_s^0$ meson, including only the $B^*K$ loop (upper panel)
and including only the $BK^*$ loop (lower panel) versus baryon density ($\rho_B/\rho_0$)
for five different values of the cutoff mass $\Lambda$.}%
\label{partbs0}%
\end{figure}

We present the in-medium $B_s$ meson mass and the mass shift
in Fig.~\ref{partbs0} as a function of baryon density ($\rho_B/\rho_0$)
including only each loop contribution, $B^*K$ or $BK^*$ loop.
(Recall that each case the $m^0$ is determined to reproduce the free space
physical $B_s$ meson mass.)
Including only the $B^*K$ loop, the mass shift 
$\Delta m_{B_s}(B^*K)$
at $\rho_0$ ranges from -74.0 to -52.7 MeV
[$m^*_{B_s}(B^*K)$ ranges from 5292.9 to 5314.3 MeV].
As for including only the $BK^*$ loop, the $B_s$ mass shift
$\Delta m_{B_s}(BK^*)$ ranges from -142.7 to -190.4 MeV
at $\rho_0$ [$m^*_{B_s}(BK^*)$ ranges 5224.2 to 5176.5 MeV].
However, for the $\Delta m_{B_s}(BK^*)$ case,
$\Lambda$ value dependence is opposite, namely
as the $\Lambda$ increases, the amount of the mass shift decreases, and the range of the changes is
less than half compared to that of $\Delta m_{B_s}(BK^*)$.
We can guess that this behavior might be related to the light $K$ meson in the loop
since the $BK$ loop in the $B_s^*$ self-energy also shows a similar behavior,
which will be shown later. Furthermore, it will be explicitly shown that the
$B^*K$ loop contribution in the total $B^*K + BK^*$ is less than 10\%; thus, the behavior of the
$B^*K$ loop contribution may be caused by some subtle mechanism or interplay of the form factor
and/or Lorentz structure in the self-energy.
We will come back to this issue later.
%%%
%%%%%%%%%%%%%%%%%%
%%%%%%%%%%%%%%%%%%%%%%

\begin{figure}[htb]%
\vspace{8ex}
%\centering
\hspace{-16ex}
\includegraphics[width=8.0cm]{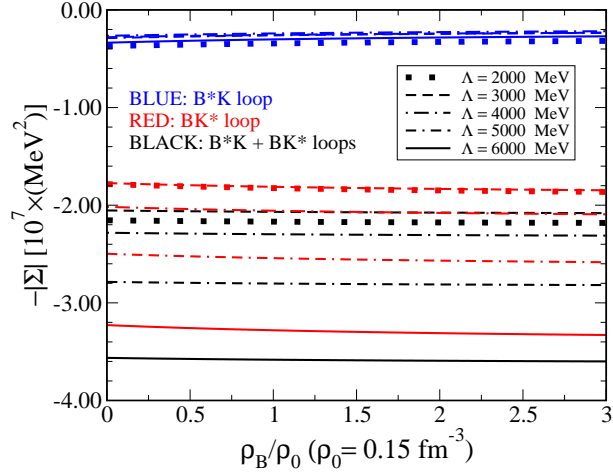}%
\caption{Decomposition of the $B^0_s$ meson total self-energy as a function of
baryon density ($\rho_B/\rho_0$)
for five different values of the cutoff mass $\Lambda$.
}%
\label{bs0se}%
\end{figure}

Next, we show in Fig.~\ref{bs0se} the decomposition of the $B_s$ total self-energy ($B^*K + BK^*$
loops) and each contribution of $B^*K$ and $BK^*$ loops as a function of nuclear matter density
($\rho_B/\rho_0$), for the same $\Lambda$ values with those used for $B_c$ and $B^*_c$.
The $B^*K$ loop gives a smaller contribution (about 10\% of the total) than the $BK^*$ heavier-pair
loop for the total self-energy.
This is again opposite to our naive expectation, but the same behavior with that for the case of
$B_c$.
We can conclude that, the lighter vector meson $K^*$ excitation (propagator)
is responsible for this result.

\begin{figure}[htb]
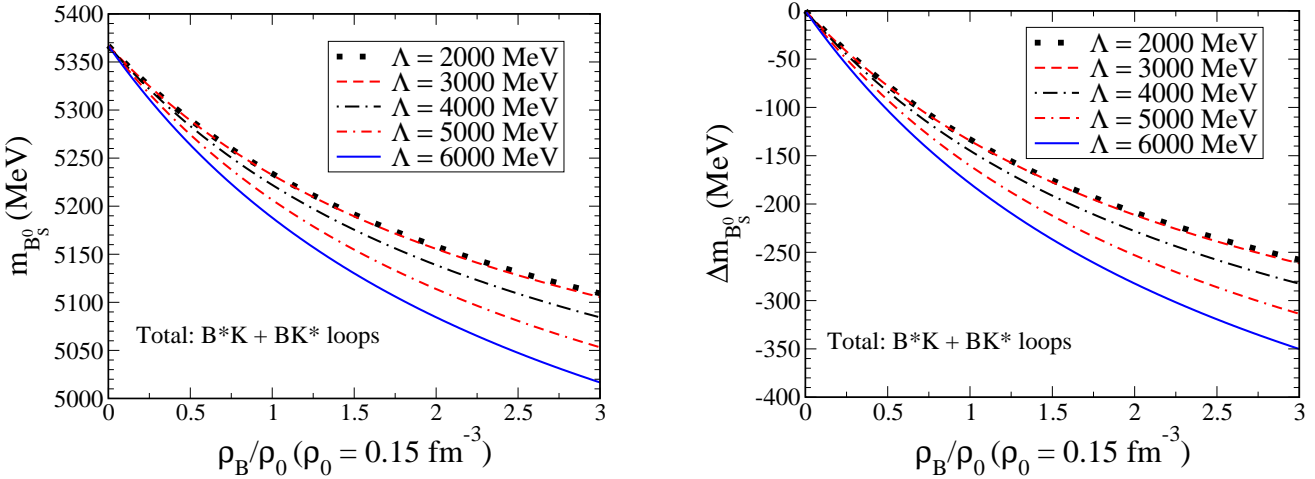

\vspace{8ex}
%\centering
\hspace{-16ex}
 \includegraphics[width=8.0cm]{Bs0_totalmass.eps}
\hspace{8ex}
 \includegraphics[width=8.0cm]{Bs0s_totalpot.eps}
 \caption{Total ($B^*K$+$BK^*$) loop contribution to the
in-medium $B^0_s$ mass (left) and mass shift (right) versus
baryon density ($\rho_B/\rho_0$) for five different
values of the cutoff mass $\Lambda$.}
\label{totbs0}
\end{figure}

Finally, the density dependence of the total mass shift
$\Delta m_{B_s}(B^*K + BK^*)$,
as well as $m^*_{B_s}(B^*K + BK^*)$ (our predictions) are shown in Fig.~\ref{totbs0}.
The cutoff $\Lambda$ value dependence for these quantities look monotonous, namely, as the $
\Lambda$ value increases their amounts increase.

This is because the $BK^*$ contribution dominates over that of the $B^*K$ loop, and
the former $\Lambda$ dependence is reflected in the total results.
At $\rho_0$, $\Delta m_{B_s}(B^*K + BK^*)$ ranges from
-133.0 to -178.8 MeV [$m^*_{B_s}(B^*K + BK^*)$ ranges from 5233.8 to 5188.0 MeV].

%%%%%%%%%%%%%%%%%%%%%%
%%%%%%%%%%%%%%%%%%%

%%%%%%%%%%%%%%%%%%
%%%%%%%%%%%%%%%%%

%%%%%%%%%%%%%%%%%%%%%%%%%%%%%%%%%%%%%%%%%%%%%%
\subsubsection{\boldmath{$B_s^*$} meson results}
%%%%%%%%%%%%%%%%%%%%%%%%%%%%%%%%%%%%%%%%%%%%%

For the $B^*_s$ self-energy, we include only the $BK$ loop.
In Table~\ref{Bssm0} we show the cutoff mass ($\Lambda$) value dependence
of the bare mass $m^0$.
Similar to the $B_s$ meson case, $\Lambda$ dependence
of $m^0$ oscillates, but the amount of oscillation is smaller.
%%%%%%%%%%%%%%%%%%%%%%%%%%%%%%%%%%%%%%%%%%%%%%%%%%%
\begin{table}[htb!]
\caption{
Cutoff mass ($\Lambda$) value dependence of $m^0$.
\label{Bssm0}
}
\begin{center}   %%%%%%%%%%%%%%%%%%%%
\begin{tabular}{c|c}
\hline
\hline
$\Lambda$ (MeV) 	&$m^0(BK)$ (MeV)     \\
\hline
\hline
2000	    & 5575.4 \\
3000      	& 5566.5 \\
4000     	& 5583.5 \\
5000      	& 5621.0 \\
6000      	& 5678.8 \\
\hline
\hline
\end{tabular}
\end{center}   %%%%%%%%%%%%%%%%%%%%%%%%%
\end{table}
%%%%%%%%%%%%%%%%%%%%%%%%%%%%%%%%%%%%%%%%%%%%%%%%%%%%

\begin{figure}[htb!]%
\vspace{8ex}
%\centering
\hspace{-16ex}
\includegraphics[width=8.0cm]{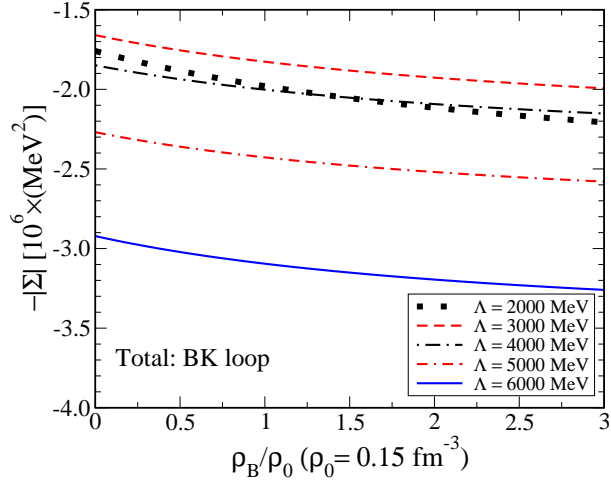}%
\caption{Total ($BK$) loop contribution for the $B^*_s$ meson self-energy
as a function of baryon density ($\rho_B/\rho_0$)
for five different values of the cutoff mass $\Lambda$.
}%
\label{bssse}%
\end{figure}

\begin{figure}[htb!]
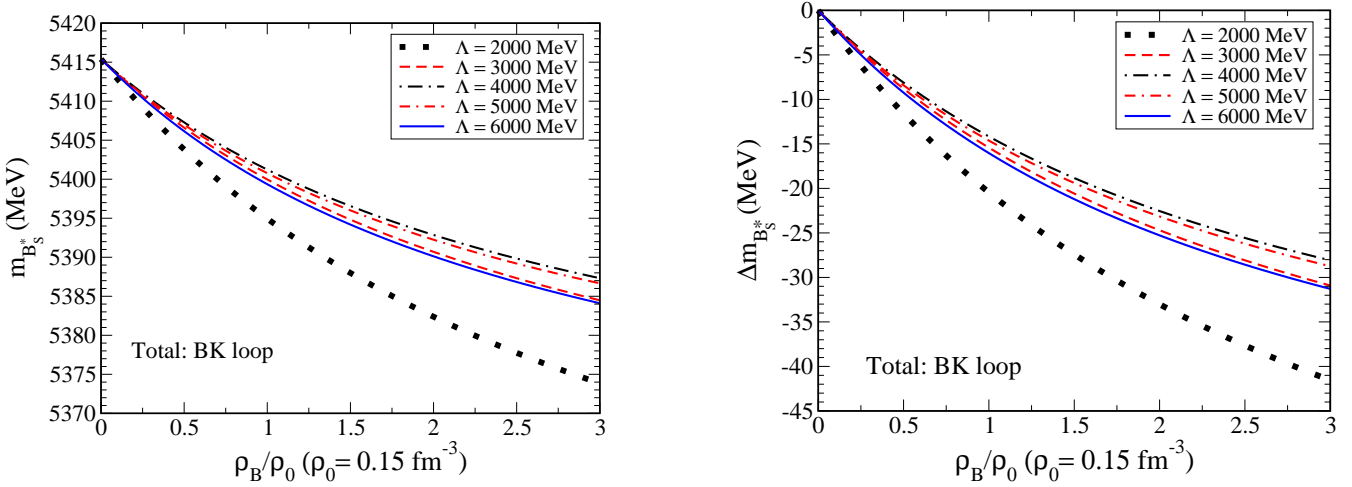

\vspace{8ex}
%\centering
\hspace{-16ex}
 \includegraphics[width=8.0cm]{Bssmass.eps}
\hspace{11ex}
 \includegraphics[width=8.0cm]{Bsspotential.eps}
 \caption{Total ($BK$) loop contribution for the
in-medium $B^*_s$ mass (left panel) and
the mass shift (right panel)
versus baryon density ($\rho_B/\rho_0$) for five different
values of the cutoff mass $\Lambda$.}
\label{totbss}
\end{figure}

The total self-energy and the mass shift (our predictions)
are presented in Figs.~\ref{bssse} and~\ref{totbss}, respectively.
The $B_s^*$ mass shift $\Delta m_{B_s^*}(BK)$ at $\rho_0$ ranges from -20.5 to -16.0 MeV
(corresponding to $m^*_{B^{*}_{s}}(BK)$ in the range from 5394.9 to 5399.4 MeV).
Similar to the $B_s$ with only the $B^*K$ loop,
the cutoff $\Lambda$ value dependence oscillates, and with the larger $\Lambda$ values,
a lesser amount of the mass shift $\Delta m_{B_s^*}(BK)$ is observed.
%%%%%%%%%%%%%%%%%%%%%%%%%%%%%
%%%%%%%%%%%%%%%%%%

%%%%%%%%%%%%%%%%%%
%%%%%%%%%%%%%%%%%

%%%%%%%%%%%%%%%%%%%%%%%%%%%%%%%%%%%%%%%%%%%%%%
\subsubsection{\boldmath{$D_s$} meson results}
%%%%%%%%%%%%%%%%%%%%%%%%%%%%%%%%%%%%%%%%%%%%%

For the $D_s$ meson self-energy, we include only the
$DK^*$ loop, only the $D^*K$ loop, and $DK^*+D^*K$ loops.
In Table~\ref{Dsm0}, we show the cutoff $\Lambda$ value dependence of $m^0$.
For the $D_s$ meson, the $\Lambda$ value dependence is monotonous
and does not show an oscillating behavior, namely, as the $\Lambda$
increases, $m^0$ increases monotonously.
%%%%%%%%%%%%%%%%%%%%%%%%%%%%%%%%%%%%%%%%%%%%%%%%%%%%%%%%%%%%%%%%%%%%
\begin{table}[htb!]
\caption{
Cutoff mass ($\Lambda$) value dependence of $m^0$ for only the $D^*K$ loop,
only the $DK^*$ loop,  and the total $D^{*}K + DK^{*}$ loops.
\label{Dsm0}
}
\begin{center}   %%%%%%%%%%%%%%%%%%%%
\begin{tabular}{c|c|c|c}
\hline
\hline
$\Lambda$ (MeV)  & $m^0(D^*K)$ (MeV) &$m^0(DK^*)$ (MeV) &$m^0(D^*K+DK^*)$ (MeV)\\
\hline
\hline
2000		& 2503.1 & 2809.9 & 3207.3 \\
3000      	& 2940.5 & 3746.5 & 4336.8 \\
4000     	& 3551.1 & 5024.7 & 5829.5 \\
5000      	& 4262.1 & 6505.5 & 7524.1 \\
6000      	& 5021.3 & 8097.4 & 9322.4 \\
\hline
\hline
\end{tabular}
\end{center}   %%%%%%%%%%%%%%%%%%%%%%%%%
\end{table}
%%%%%%%%%%%%%%%%%%%%%%%%%%%%%%%%%%%%%%%%%%%%%%%%%%%%%%%%%%%%%%%%%

\begin{figure}[htb!]
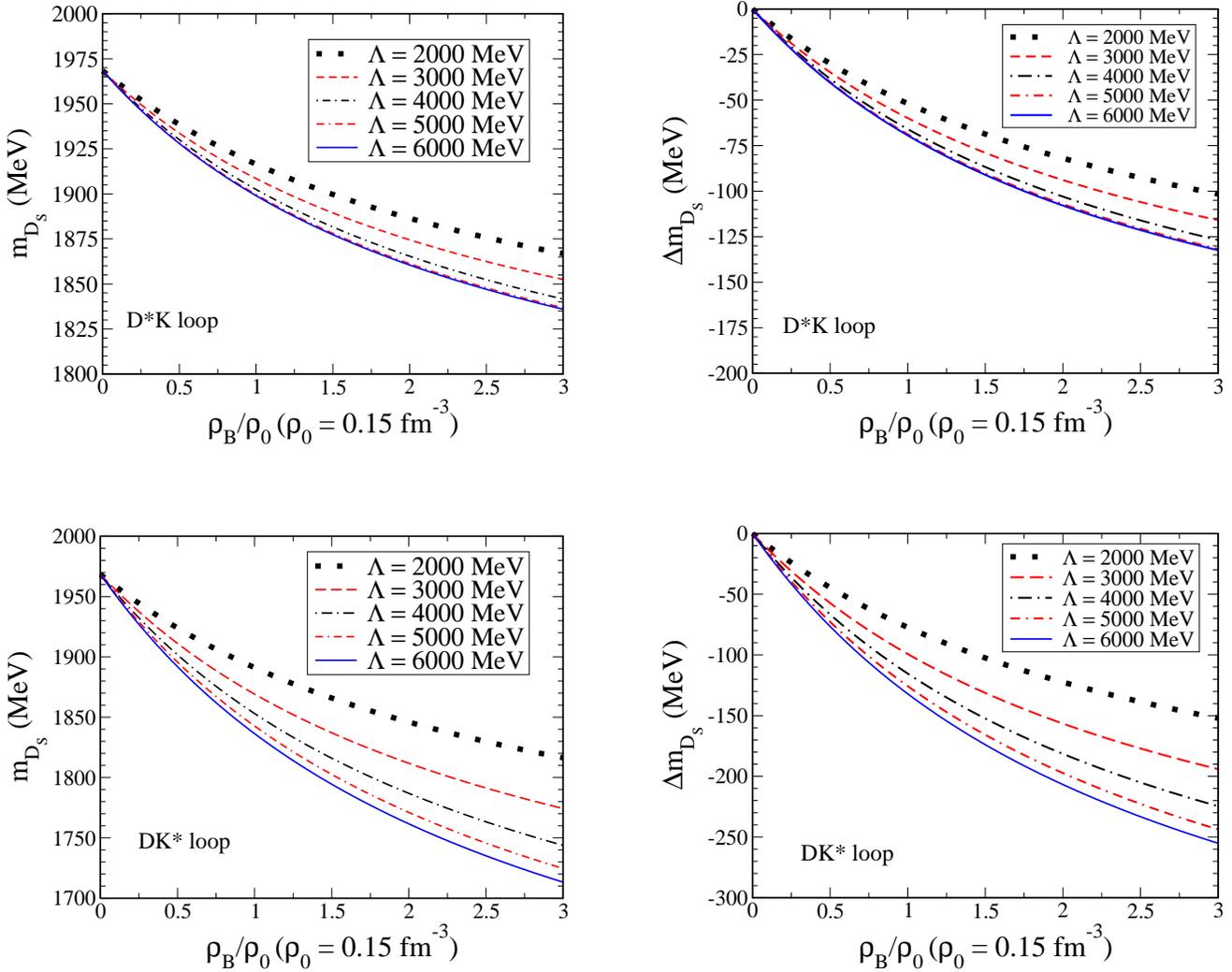
%
\vspace{8ex}
%\centering
\hspace{-16ex}
\includegraphics[width=8.0cm]{Dcs_DsK_mass.eps}
\hspace{8ex}
\includegraphics[width=8.0cm]{Dcs_DsK_pot.eps}
\\
\vspace{8ex}
\hspace{-16ex}
\includegraphics[width=8.0cm]{Dcs_DKs_mass.eps}
\hspace{8ex}
\includegraphics[width=8.0cm]{Dcs_DKs_pot.eps}
\caption{Density dependence of the $D_s$ meson in-medium mass (left panel)
and the mass shift (right panel), including only the $D^*K$ loop (upper panel)
and including only the $DK^*$ loop (lower panel)
for five different values of cutoff mass $\Lambda$.
}%
\label{partds}%
\end{figure}

In Fig.~\ref{partds}, we present the in-medium $D_s$ meson mass (left panel)
and mass shift (right panel) in symmetric nuclear matter
as a function of baryon density ($\rho_B/\rho_0$)
including only each loop contribution, only the $D^*K$ loop (upper panel),
and only the $DK^*$ loop (lower panel).
%%%%%%%%%%%%%%%%%%
%%%%%%%%%%%%%%%%%%%%%%%%%%

The value of the mass shift for only the $D^*K$ loop  $\Delta m_{D_s}(D^*K)$
ranges from -51.9 to -69.4 MeV [$m^*_{D_s}(D^*K)$ from 1916.5 to 1899.0 MeV],
while the inclusion of only the $DK^*$ loop 
$\Delta m_{D_s}(DK^*)$
gives mass shift at $\rho_0$ ranges from -77.3 to -132.0 MeV [$m^*_{D_s}(DK^*)$ from 1891.1
to 1836.3 MeV].

\begin{figure}[htb!]%
\vspace{8ex}
%\centering
\hspace{-16ex}
\includegraphics[width=8.0cm]{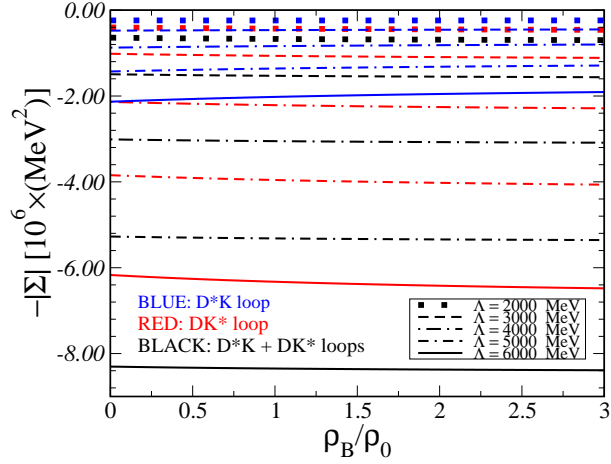}%
\caption{Partial and total loop contributions to the $D_s$ meson self-energy as a function of
baryon density for five different values of the cutoff mass $\Lambda$.
}%
\label{dsse}%
\end{figure}

\begin{figure}[htb!]
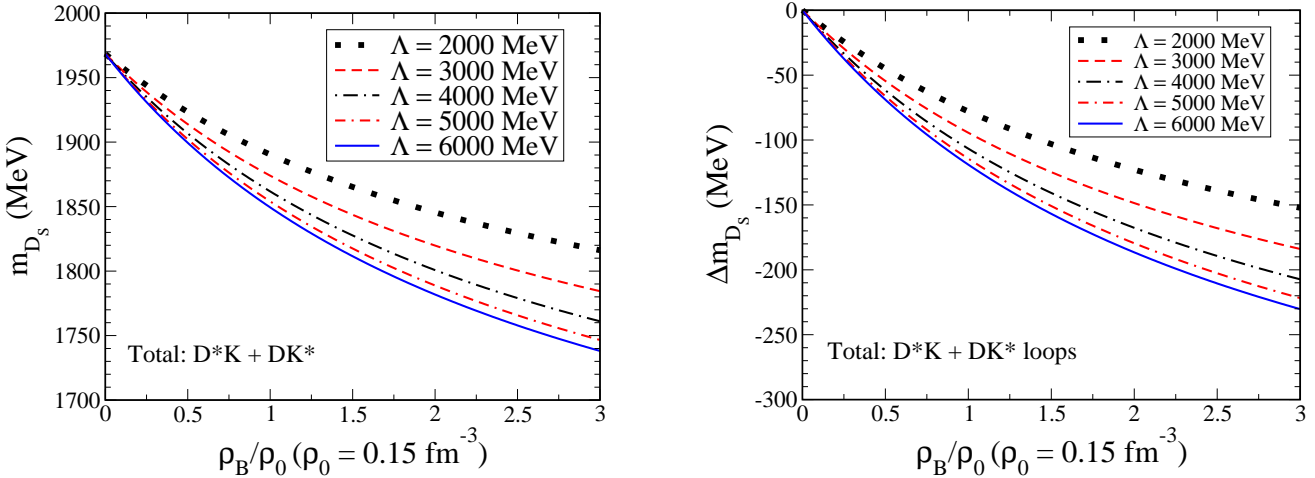

\vspace{8ex}
%\centering
\hspace{-16ex}
 \includegraphics[width=8.0cm]{Dcs_totalmass.eps}
\hspace{8ex}
 \includegraphics[width=8.0cm]{Dcs_totalpot.eps}
 \caption{Total ($D^*K$ + $DK^*$) loop contribution to the
in-medium $D_s$ mass (left) and mass shift (right) versus baryon density for five different
values of the cutoff mass $\Lambda$.}
 \label{totds}
\end{figure}

The predicted results for the $D_s$ meson are presented in Figs.~\ref{dsse}
and~\ref{totds}, respectively, for the decomposition of the $D_s$ self-energy,
and the total $D^*K+DK^*$ contribution for the in-medium mass (left panel)
and the mass shift (right panel).
It is evident from Fig.~\ref{dsse} that the $DK^*$ loop gives the dominant contribution
for the mass shift. Again this is opposite from our naive expectation based
on the total mass value comparison of Eq.~(\ref{tmass3}) and shows the same
conclusion that the lighter vector meson $K^*$ excitation in the intermediate state
gives the larger contribution than that of the $D^*$.
%%%%%%%%%%%%%%%%%%%%%%%%%%
%%%%%%%%%%%%%%%%%%%

%%%%%%%%%%%%%%%%%%%
%%%%%%%%%%%%%%%%%

This conclusion agrees with the $B_c$ and $B_s$ cases
and confirms that the corresponding lighter vector mesons $D^*$ and $K^*$
meson excitations give the dominant contributions for the self-energies.

The total, $DK^* + D^*K$ loop contribution (our prediction),
gives the mass shift $\Delta m_{D_s}(DK^* + D^*K)$
at $\rho_0$ ranging from -78.0 to -119.0 MeV [corresponding in-medium mass of
$m^*_{D_s}(DK^* + D^*K)$ from 1890.4 to 1849.4 MeV].

%%%%%%%%%%%%%%%%%%%%%%%%%%%%%%%%%%%%%%%%%%%%%%
\subsubsection{\boldmath{$D_s^*$} meson results}
%%%%%%%%%%%%%%%%%%%%%%%%%%%%%%%%%%%%%%%%%%%%%

Finally, we discuss he $D^*_s$ meson, including only the $DK$ loop.
In Table~\ref{Dssm0}, we show the cutoff mass ($\Lambda$) value dependence of $m^0$.
In this case as well, the $\Lambda$ dependence is monotonous, without oscillation.
%%%%%%%%%%%%%%%%%%%%%%%%%%%%%%%%%%%%%%%%%%%%%%
\begin{table}[htb!]
\caption{
Cutoff $\Lambda$ dependence of $m^0$ for the $D_s^*$ meson.
\label{Dssm0}
}
\begin{center}   %%%%%%%%%%%%%%%%%%%%
\begin{tabular}{c|c}
\hline
\hline
$\Lambda$ (MeV) &$m^0(DK)$ (MeV)   \\
\hline
\hline
2000	    & 2198.7 \\
3000      	& 2288.6 \\
4000     	& 2441.2 \\
5000      	& 2656.0 \\
6000      	& 2924.4 \\
\hline
\hline
\end{tabular}
\end{center}   %%%%%%%%%%%%%%%%%%%%%%%%%
\end{table}
%%%%%%%%%%%%%%%%%%%%%%%%%%%%%%%%%%%%%%%%%%%%%%

\begin{figure}[htb!]%
\vspace{8ex}
%\centering
\hspace{-16ex}
\includegraphics[width=8.0cm]{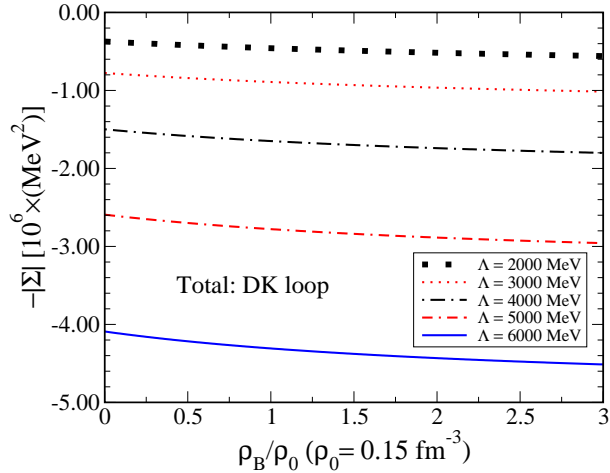}%
\caption{Baryon density ($\rho_B/\rho_0$) dependence of
the $D^*_s$ meson self-energy ($DK$ loop (total)) for five different values of the
cutoff mass $\Lambda$.
}%
\label{dssse}%
\end{figure}

\begin{figure}[htb!]
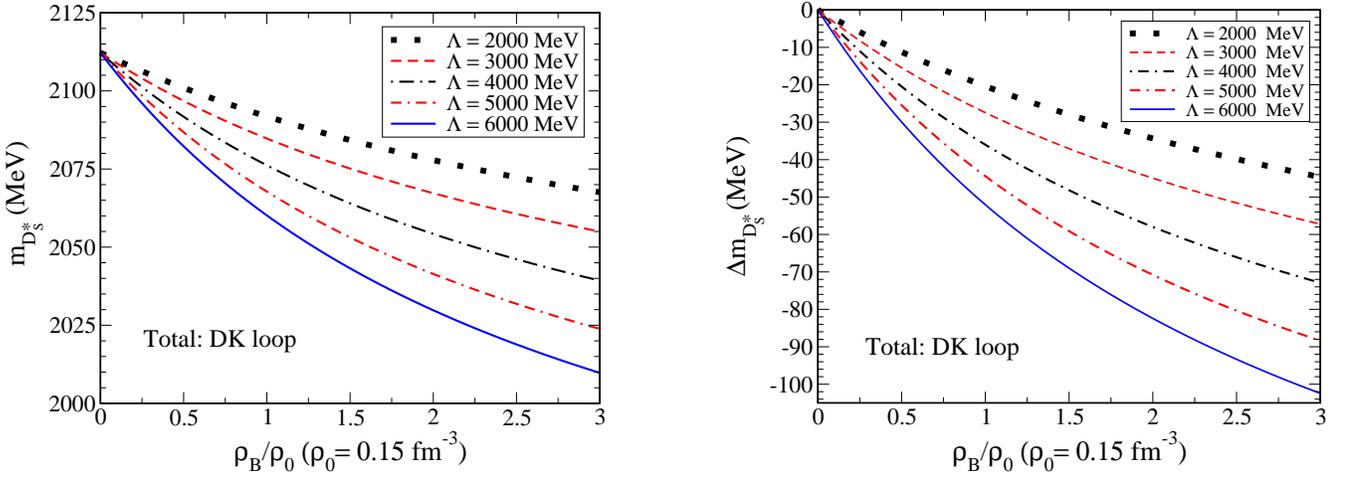

\vspace{8ex}
%\centering
\hspace{-16ex}
\includegraphics[width=8.0cm]{Dcssmass.eps}
\hspace{10ex}
\includegraphics[width=8.0cm]{Dcsspot.eps}
\caption{Baryon density ($\rho_B/\rho_0$) dependence of the $D_s^*$ in-medium mass (left panel) and
mass shift (right panel) for five different values of the cutoff mass $\Lambda$
}
\label{totdss}
\end{figure}

Density dependence of the $D_s^*$ meson mass shift and in-medium mass
are, respectively, presented in Figs.~\ref{dssse} (left panel)
and~\ref{totdss} (right panel) for
five values of cutoff $\Lambda$.
%%%%%%%%%%%%%%%%%%%%%
%%%%%%%%%%%%%%%%%%

%%%%%%%%%%%%%%%%%%
%%%%%%%%%%%%%%%%%

The cutoff mass ($\Lambda$) value dependence of the $D_s^*$ mass shift and in-medium mass
is monotonous, namely, as the $\Lambda$ value increases, the amount of the
mass shift increases.
The $D^*_s$ meson mass shift $\Delta m_{D_s^*}(DK)$
at $\rho_0$ ranges from -20.2 to -52.1 MeV
[corresponding to the in-medium mass $m^*_{D_s^*}(DK)$
ranges from 2091.8 to 2060.1 MeV].

%%%%%%%%%%%%%%%%%%%%%%%%%%%%%%%%%%%%%%%%%%%%%%
\subsubsection{\boldmath{$D_s$ and $D_s^*$} results for SU(4) coupling}
%%%%%%%%%%%%%%%%%%%%%%%%%%%%%%%%%%%%%%%%%%%%%
To make a comparison in the same spirit as that of Sec.~\ref{hvqrkcomp},
we now calculate the mass shift of the $D_s$ and $D^*_s$ mesons using the coupling constant
extracted from the universal coupling $g$ in the SU(4) sector,
originally obtained from $g_{J/\Psi DD}$ in Eq.~(\ref{gJPsi}). We get the coupling constants for
the vertices $D_s D^* K$, $D_s D K^*$, and
$D^*_s D K$ by the relation 

\begin{equation}
g_{D_{s} KD^*} = g_{D_{s} DK^*} = g_{D^{*}_{s} DK} = 
\frac{g}{2\sqrt{2}} \approx 6.7.  
\end{equation}

\begin{figure}[htb!]
\vspace{8ex}
%\centering
\hspace{-16ex}
\includegraphics[width=8.0cm]{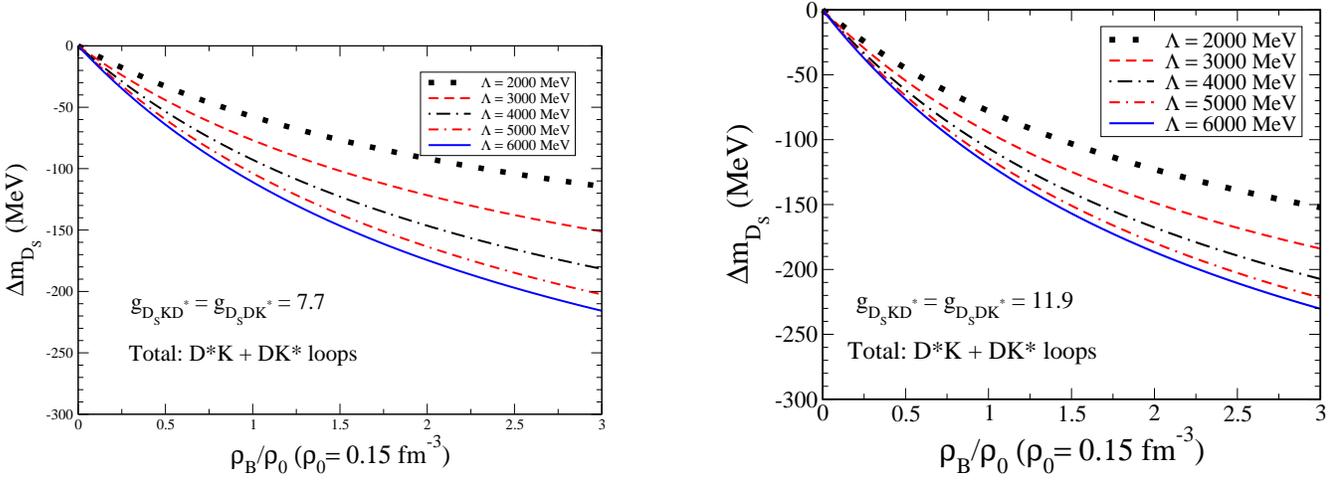}
\hspace{10ex}
\includegraphics[width=8.0cm]{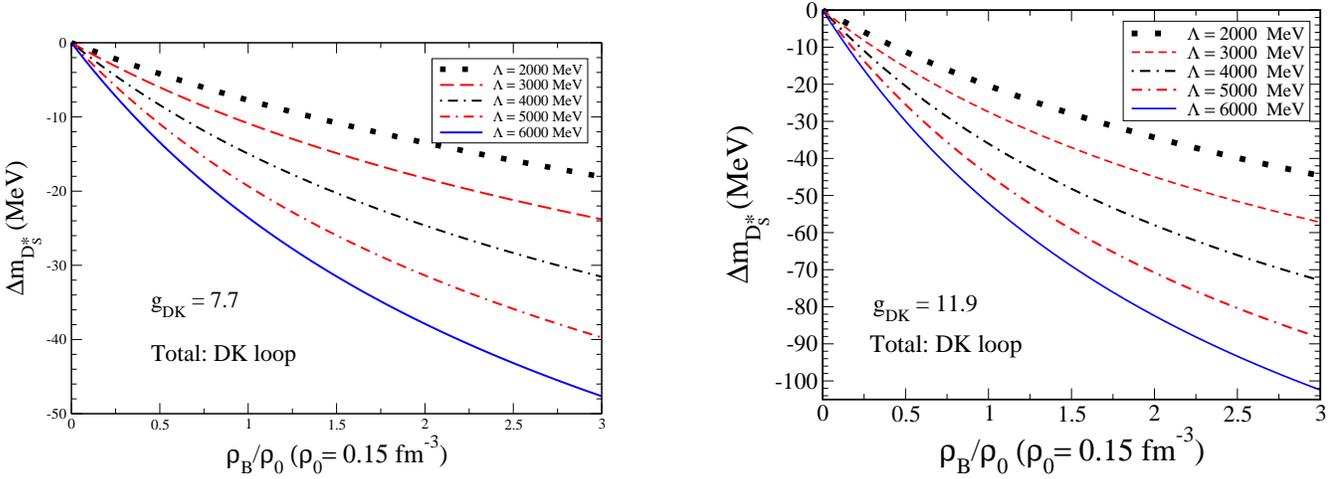}
\caption{Comparison of the in-medium mass shift of $D_s$ for $g = 7.7$ (left) and for $g = 11.9$ (right).
}
\label{su45ds}
\end{figure}

The values for the coupling constants in SU(4) and SU(5) are summarized in Table~\ref{tblcplDs}, and
the results for the mass shifts calculated with those couplings are presented in Figs.~\ref{su45ds}
and~\ref{su45dss}.

At normal nuclear matter density, the mass shifts of $D_s$ and $D^*_s$, calculated with
$g$ in SU(4) symmetry
and SU(5) symmetry, are
$\Delta m_{D_s}$($g$SU(4)) = -57.6 to -111.0, $\Delta m_{D_s}$($g$SU(5)) = -78.0 to -119.0, 
$\Delta m_{D^*_s}$($g$SU(4)) = -7.7 to -23.6, and $\Delta m_{D^*_s}$($g$SU(5)) = -20.2 to -52.1. 
As one can expect, the larger value of $g$, obtained in the SU(5) symmetry, makes the mass shift of
the mesons more negative in comparison to the results obtained with $g$ in SU(4). The
absolute value of the mass shift of $D_s$ and $D^*_s$ are smaller in SU(4) symmetry.
However, the inequality relations among the mass shift of
$B_c$, $B^*_c$, $B^0_s$, and $B^*_s$ remained the same as summarized next.
\\

\noindent
$Summary:$\\

$|\Delta m_{B_c}| < |\Delta m_{B^0_s}| > |\Delta m_{D_s}|$

$|\Delta m_{B^*_c}| \approx |\Delta m_{B^{*}_s}| \lesssim |\Delta m_{D^*_s}|$

%%%%%%%%%%%%%%%%%%%%%%%%%%%%%%%%%%%%%%%%%%%%%%%%%%%%%%%%%%%%%%%%%%%%%%%
\begin{table}
\caption{\label{tblcplDs} Coupling constant values in SU(4) and SU(5) symmetries.}
\begin{center}
%\scalebox{0.8}{
\begin{tabular}{ll|r|r}
  \hline \hline
%  & & \multicolumn{2}{c}{Coupling constants in SU(4) and SU(5)} \\
%\hline
& & SU(4) & SU(5) \\
\hline
&$g$ & 18.9 & 33.4 \\
& $g_{D_{s}D^{*}K}$ & 6.7 & 11.9 \\
& $g_{D^{*}_{s}DK}$ & 6.7 & 11.9 \\
\hline \hline
\end{tabular}
%}
\end{center}
\end{table}
%
%%%%%%%%%%%%%%%%%%%%%%%%%%%%%%%%%%%%%%%%%%%%%%%%%%%%%%%%%%%%%%%%%%%%%%

%%%%%%%%%%%%%%%%%%
%%%%%%%%%%%%%%%%%

%%%%%%%%%%%%%%%%%%
\begin{figure}[htb!]
\vspace{8ex}
%\centering
\hspace{-16ex}
\includegraphics[width=8.0cm]{Dcsspot_gSU4.eps}
\hspace{10ex}
\includegraphics[width=8.0cm]{Dcsspot_gSU5.eps}
\caption{Comparison of the in-medium mass shift of $D^*_s$ for $g = 7.7$ (left) and for $g = 11.9$ (right).
}
\label{su45dss}
\end{figure}
%%%%%%%%%%%%%%%%%

%%%%%%%%%%%%%%%%%%%%%%%%%%%%%%%%%%
\section{Summary and Conclusions}
\label{seconcl}
%%%%%%%%%%%%%%%%%%%%%%%%%%%%%%%%%

We have estimated for the first time the in-medium mass shift of two-flavored heavy mesons, $B_c,
B_c^*, B_s, B_s^*, D_s$, and $D_s^*$ in symmetric nuclear matter.
The estimates are made by evaluating the lowest order one-loop self-energies of these mesons at the
hadronic level.
The results show the negative mass shift for all these mesons.

We have derived and used SU(5) effective Lagrangians for
the interactions of $B_c B^*D$, $B_c BD^*$, $B^*_c BD$, $B_s B^*K$,
$B_s BK^*$, $B^*_s BK$, $D_s D^*K$, $D_s DK^*$, and $D^*_s DK$.

In symmetric nuclear matter, the in-medium masses of the intermediate state
excited mesons appearing in the self-energy loops
are calculated by the quark-meson coupling model.
The enhancement of the self-energy in symmetric nuclear matter results in the negative mass shift
of the two-flavored heavy mesons.
For the estimates, we use the form factors to regularize the divergent loop integrals with five
values of cutoff mass values.

Our predictions (total loop contributions)
for the mass shift ranges of these mesons at nuclear matter saturation density ($\rho_0=0.15$
fm$^{-3}$) with five different cutoff mass values in form factors are
(1) for $B_c$, [-90.4, -101.1] MeV,
(2) for $B^*_c$, [-14.5, -19.7] MeV,
(3) for $B_s$, [-133.0, -178.8] MeV,
(4) for $B^*_s$, [-20.5, -16.0] MeV,
(5) for $D_s$, [-78.0, -119.0] MeV,
and
(6) for $D^*_s$, [-20.2, -52.1] MeV.

We have also compared the mass shift of $B_c$, and $B_c^*$ mesons,
respectively, with those of pseudoscalar quarkonia ($\eta_b$ and $\eta_c$),
and vector quarkonia ($\Upsilon$ and $J/\psi$).
For the $B_c$ meson, the results show that
(i) the $B_c$ in-medium mass shift amount is larger than
those of the $\eta_b$ and $\eta_c$.
This is different from our naive expectation that the $B_c$ mass shift would be in between those of
the $\eta_b$ and $\eta_c$.
(ii) The heavier meson pair $BD^*$ loop contribution is more dominant than that of the lighter meson
$B^*D$ loop.
This is because for the former, the lighter vector meson $D^*$ contributes more
effectively than that of the $B^*$ due to the spin-1 propagator Lorentz structure.
Similar things are also observed for the other pseudoscalar meson cases,
$B_s$ and $D_s$, namely, the loop with the lighter vector meson excitation
gives dominant contribution: $BK^*$ loop for the $B_s$ mass shift
and $DK^*$ loop for the $D_s$ mass shift.

While for the $B_c^*$ meson, the mass shift result shows that it is overall in between those of $\Upsilon$ and $J/\psi$.
In addition, the amount of the in-medium mass shift is smaller than that of the $B_c$.
Similar things hold for the $B_s^*$ and $D_s^*$ meson cases.
Namely, $B_s^*$ mass shift amount is smaller than that of the $B_s$, and $D_s^*$ mass shift amount is smaller than that of the $D_s$.

We have not included possible in-medium widths of any mesons appearing in the calculations.
However, we expect that the mass shift (scalar potentials) for the mesons
studied are strong enough to bind these mesons to atomic nuclei.
The study of meson-nucleus bound states for these mesons including
the Coulomb potentials and/or the in-medium widths,
will be performed in the near future.

%%%%%%%%%%%%%%%%%%%%%%%%%%%%%%%%%%%%%%%%
\begin{acknowledgments}
G.N.Z and K.T. acknowledge the support and warm hospitality of APCTP
(Asia Pacific Center for Theoretical Physics) during the Workshop (APCTP PROGRAMS 2023)
“Origin of Matter and Masses in the Universe: Hadrons in free space, dense nuclear medium,
and compact stars,'' where important discussions and development were achieved
on the topic.
G.N.Z and K.T. also thank the OMEG (Origin of Matter and Evolution of Galaxies) Institute at
Soongsil University for the support in many aspects during the collaboration visit in Korea.
G.N.Z~and S.L.P.G.B.~were supported by the Coordena\c{c}\~ao de Aperfei\c{c}oamento de Pessoal de
N\'ivel Superior- Brazil (CAPES).
K.T.~was supported by Conselho Nacional de Desenvolvimento
Cient\'{i}fico e Tecnol\'ogico (CNPq, Brazil), Processes No.~313063/2018-4,
No.~426150/2018-0, No.~304199/2022-2,
and FAPESP Process No.~2019/00763-0 and No.~2023/07313-6.
The work of authors was in the projects of
Instituto Nacional de Ci\^{e}ncia e
Tecnologia - Nuclear Physics and Applications
(INCT-FNA), Brazil, Process No.~464898/2014-5.
\end{acknowledgments}
%%%%%%%%%%%%%%%%%%%%%%%%%%%%%%%%%%%%%%%

%%%%%%%%%%%%%%%%%%%%%%%%%%%%%%%%%%%%%%%%%%%%%%%%%%%%%%%%%%%%%%%%%%%

%%%%%%%%%%%%%%%%%%%%%

%%%%%%%%%%%%%%%%%%%%%%%%%%%%%%%%%%%%%%%%%%%%%%%%%%%%%%%%%%%%%%%%%%%%%%%%%%%%%%%%%%%%%%%
\end{document}